\documentclass[10]{article}

\usepackage{psfig,labelfig} 
\usepackage{cite}				
\usepackage{amsmath}    
\usepackage{graphicx}   
\usepackage{subfigure}	
\usepackage{color}      
\usepackage{hyperref}   
\usepackage{amssymb}    
\usepackage{cancel}			
\usepackage{setspace}		

\def\d{\partial}				
\def\E{\mathbb{E}}			
\def\P{\mathbb{P}}			
\def\R{\mathbb{R}}			
\def\L{{\cal L}}				

\def\A{{\cal A}}

\def\M{{\cal M}}

\def\<{\left\langle} 
\def\>{\right\rangle}
\def\nn{\nonumber}
\def\fhat{\widehat{f}}

\def\Ghat{\widehat{G}}
\def\hhat{\widehat{h}}
\def\Lhat{\widehat{\L}}
\def\Ahat{\widehat{\A}}
\def\Phat{\widehat{P}}

\def\eps{\epsilon}

\usepackage{amsthm}																
\theoremstyle{plain}
\newtheorem{theorem}{Theorem}
\newtheorem{lemma}[theorem]{Lemma}								

\theoremstyle{definition}

\theoremstyle{remark}

\numberwithin{equation}{section}	
\numberwithin{theorem}{section}

\numberwithin{equation}{section}				

\title{A Fast Mean-Reverting Correction to Heston's Stochastic Volatility Model}
\author{Jean-Pierre Fouque\thanks{Department of Statistics \& Applied Probability,
 University of California,
        Santa Barbara, CA 93106-3110, {\em fouque@pstat.ucsb.edu}. Work partially supported by NSF grant DMS-0806461.} \and Matthew J. Lorig\thanks{Department of Statistics \& Applied Probability,
 University of California,
        Santa Barbara, CA 93106-3110, {\em lorig@pstat.ucsb.edu}.}}
        \date{\today}
        
\begin{document}
\maketitle

\begin{abstract}
We propose a multi-scale stochastic volatility model in which a fast mean-reverting factor of volatility is built on top of the Heston stochastic volatility model.  A singular pertubative expansion is then used to obtain an approximation for European option prices.  The resulting pricing formulas are semi-analytic, in the sense that they can be expressed as integrals.  Difficulties associated with the numerical evaluation of these integrals are discussed,  and techniques for avoiding these difficulties are provided. Overall, it is shown that computational complexity for our model is comparable to the case of a pure Heston model, but our correction brings significant flexibility in terms of fitting to the implied volatility surface. This is illustrated numerically and with option data.
\end{abstract}

%


\section{Introduction}
Since its publication in 1993, the Heston model \cite{heston} has received considerable attention from academics and practitioners alike.  The Heston model belongs to a class of models known as stochastic volatility models.  Such models relax the assumption of constant volatility in the stock price process, and instead, allow volatility to evolve stochastically through time.  As a result, stochastic volatility models are able to capture some of the well-known features of the implied volatility surface, such as the volatility smile and skew (slope at the money).  Among stochastic volatility models, the Heston model enjoys wide popularity because it provides an explicit, easy-to-compute, integral formula for calculating European option prices.  In terms of the computational resources needed to calibrate a model to market data, the existence of such a formula makes the Heston model extremely efficient compared to models that rely on Monte Carlo techniques for computation and calibration.

Yet, despite its success, the Heston model has a number of documented short-comings.  For example, it has been statistically verified that the model misprices far in-the-money and out-of-the-money European options \cite{fiorentini2002}, \cite{zhang}.  In addition, the model is unable to simultaneously fit implied volatility levels across the full spectrum of option expirations available on the market \cite{gatheral2}.  In particular, the Heston model has difficulty fitting implied volatility levels for options with short expirations \cite{gatheral}.  In fact, such problems are not limited to the Heston model.  Any stochastic volatility model in which the volatility is modeled as a one-factor diffusion (as is the case in the Heston model) has trouble fitting implied volatility levels across all strikes and maturities \cite{gatheral}.

One possible explanation for why such models are unable to fit the implied volatility surface is that a single factor of volatility, running on a single time scale, is simply not sufficient for describing the dynamics of the volatility process.  Indeed, the existence of several stochastic volatility factors running on different time scales has been well-documented in literature that uses empirical return data \cite{alizadeh2001}, \cite{anderson}, \cite{chernov2003}, \cite{engle}, \cite{fouque2}, \cite{hillebrand}, \cite{lebaron}, \cite{melino}, \cite{muller}.  Such evidence has led to the development of multi-scale stochastic volatility models, in which instantaneous volatility levels are controlled by multiple diffusions running of different time scales (see, for example, \cite{fouque}).  We see value in this line of reasoning and thus, develop our model accordingly. 

Multi-scale stochastic volatility models represent a struggle between two opposing forces.  On one hand, adding a second factor of volatility can greatly improve a model's fit to the implied volatility surface of the market.  On the other hand, adding a second factor of volatility often results in the loss of some, if not all, analytic tractability.  Thus, in developing a multi-scale stochastic volatility model, one seeks to model market dynamics as accurately as possible, while at the same time retaining a certain level of analyticity.  Because the Heston model provides explicit integral formulas for calculating European option prices, it is an ideal template on which to build a multi-scale model and accomplish this delicate balancing act.

In this paper, we show one way to bring the Heston model into the realm of multi-scale stochastic volatility models without sacrificing analytic tractability.  Specifically, we add a fast mean-reverting component of volatility on top of the Cox--Ingersoll--Ross (CIR) process that drives the volatility in the Heston model.  Using the multi-scale model, we perform a singular perturbation expansion, as outlined in \cite{fouque}, in order to obtain a correction to the Heston price of a European option.  This correction is easy to implement, as it has an integral representation that is quite similar to that of the European option pricing formula produced by the Heston model.

The  paper is organized as follows.  In Section \ref{sec:multiscale} we introduce the multi-scale stochastic volatility model and  we derive the resulting pricing partial differential equation (PDE) and boundary condition for the European option pricing problem.  In Section \ref{sec:asmptotics} we use a singular perturbative expansion to derive a PDE for a correction to the Heston price of a European option and in Section \ref{sec:P0P1} we obtain a solution for this PDE.  A proof of the accuracy of the pricing approximation is provided in Section \ref{sec:accuracy}.   In Section \ref{sec:impliedvolatility} we examine how the implied volatility surface, as obtained from the multi-scale model, compares with that of the Heston model, and in Section \ref{sec:data} we present an example of calibration to market data.    In Appendix \ref{sec:heston} we review the dynamics of the Heston Stochastic volatility model under the risk-neutral measure, and present the pricing formula for European options.  An explicit formula for the correction is given in Appendix \ref{sec:P1details}, and the issues associated with numerically evaluating the integrals-representations of option prices obtained from the multi-scale model are explored in Appendix \ref{sec:numerical}.

\section{Multi-Scale Model and Pricing PDE} \label{sec:multiscale}
Consider the price $X_t$ of an asset (stock, index, ...) whose dynamics under the pricing risk-neutral measure is described by the following system of stochastic differential equations:
\begin{eqnarray}
	dX_t &=&	r X_t dt + \Sigma_t X_t dW_t^x \label{eq:X}, \\
	\Sigma_t &=& \sqrt{Z_t}\, f(Y_t)	\label{eq:Sigma} ,\\
	dY_t &=&	\frac{Z_t }{\eps} (m - Y_t) dt +\nu \sqrt{2} \sqrt{\frac{Z_t}{\eps}}\,  dW_t^y \label{eq:OU} ,\\
	dZ_t &=& \kappa (\theta - Z_t) dt + \sigma \sqrt{Z_t}\, dW_t^z \label{eq:CIR}.
	\end{eqnarray}
Here, $W_t^x$, $W_t^y$ and $W_t^z$ are one-dimensional Brownian motions with the correlation structure
\begin{eqnarray}
d\<W^x,W^y\>_t &=& \rho_{xy} dt \label{eq:rhoXY}, \\
	d\<W^x,W^z\>_t &=& \rho_{xz} dt \label{eq:rhoXZ}, \\
	d\<W^y,W^z\>_t &=& \rho_{yz} dt \label{eq:rhoYZ},
\end{eqnarray}
where the correlation coefficients $\rho_{xy}$ $\rho_{xz}$ and $\rho_{yz}$ are constants satisfying $\rho_{xy}^2<1, \rho_{xz}^2<1, \rho_{yz}^2<1$, and $\rho_{xy}^2+\rho_{xz}^2+\rho_{yz}^2-2\rho_{xy}\rho_{xz}\rho_{yz}<1$ in order to ensure positive definiteness of the covariance matrix of the three Brownian motions.

As it should be, in (\ref{eq:X})  the  stock price discounted by the risk-free rate $r$  is a martingale under the pricing risk neutral measure.
 The volatility $\Sigma_t$ is driven by two processes $Y_t$ and $Z_t$,  through the product $\sqrt{Z_t}\, f(Y_t)$.  The process $Z_t$ is a Cox--Ingersoll--Ross (CIR) process with long-run mean $\theta$, rate of mean reversion $\kappa$, and ``CIR-volatility'' $\sigma$.  We assume that $\kappa$, $\theta$ and $\sigma$ are positive, and that $2 \kappa \theta \geq \sigma^2$, which ensures that $Z_t>0$ at all times, under the condition $Z_0>0$.
\par
Note that given $Z_t$, the process $Y_t$ in (\ref{eq:OU}) appears as an Ornstein--Uhlenbeck (OU) process evolving on the time scale $\eps/Z_t$, and with the invariant (or long-run) distribution ${\cal N}(m,\nu^2)$. This way of ``modulating" the rate of mean reversion of the process $Y_t$ by $Z_t$ has also been used in \cite{cfps} in the context of interest rate modeling. 

Multiple time scales are incorporated in this model through the parameter $\eps>0$, which is intended to be  small, so that $Y_t$ is fast-reverting. 
\par
We do not specify the precise form of $f(y)$ which will not play an essential role in the asymptotic results derived in this paper.  However, in order to ensure $\Sigma_t$ has the same behavior at zero and infinity as in the case of a pure Heston model, we assume there exist constants $c_1$ and $c_2$ such that $0<c_1 \leq f(y) \leq c_2 < \infty$ for all  $y \in \R$. Likewise, the particular choice of an OU-like process for $Y_t$ is not crucial in the analysis. The mean-reversion aspect (or ergodicity)  is the important property. In fact, we could have chosen  $Y_t$ to be a CIR-like process instead of an OU-like process without changing the nature of the correction to the Heston model presented in the paper.

Here, we consider the unique strong solution to (\ref{eq:X}--\ref{eq:CIR}) for a fixed parameter $\eps>0$. Existence and uniqueness is easily obtained by (i) using the classical existence and uniqueness result for the CIR process $Z_t$ defined by (\ref{eq:CIR}), (ii) using the representation  (\ref{eq:Yexplicit}) of the process $Y_t$ to derive moments for  a fixed $\eps>0$, (iii) using the exponential formula for $X_t$:
\[
X_t=x\exp\left(\int_0^t\left(r-\frac{1}{2}\Sigma_s^2\right)ds+\int_0^t\Sigma_sdW^x_s\right).
\]

\par
We note that if one chooses $f(y)=1$, the multi-scale model becomes $\eps$-independent and reduces to the pure Heston model expressed under the risk-neutral measure with stock price $X_t$ and stochastic variance $Z_t$:
\begin{eqnarray*}
	dX_t &=&	r X_t dt + \sqrt{Z_t} X_t dW_t^x ,\\
	dZ_t &=& \kappa (\theta - Z_t) dt + \sigma \sqrt{Z_t}\, dW_t^z. \\
	d\<W^x,W^z\>_t &=& \rho_{xz} dt.
\end{eqnarray*}
 Thus, the multi-scale model can be thought of as a Heston-like model with a fast-varying factor of volatility, $f(Y_t)$, build on top of the CIR process $Z_t$, which drives the volatility in the Heston Model.

We consider a European option expiring at time $T>t$ with payoff $h(X_T)$.  As the dynamics of the stock in the multi-scale model are specified under the risk-neutral measure, the price of the option, denoted by $P_t$, can be expressed as an expectation of the option payoff, discounted at the risk-free rate:
\begin{eqnarray*}
	P_t= \E  \left. \left[ e^{-r(T-t)} h(X_T) \right| X_t , Y_t , Z_t  \right]=: P^\eps(t,X_t,Y_t,Z_t),
\end{eqnarray*} 
where we have used the Markov property of  $(X_t,Y_t,Z_t)$, and defined the pricing function $P^\eps(t,x,y,z)$, the superscript $\eps$  denoting the dependence  on the small parameter $\eps$.  Using the Feynman--Kac formula,  $P^\eps(t,x,y,z)$ satisfies the following PDE and boundary condition:
\begin{eqnarray}
	\L^\eps P^\eps(t,x,y,z) &=& 0 , \label{eq:PepsPDE} \\
	\L^\eps &=& \frac{\d}{\d t} + \L_{(X,Y,Z)} - r\, , \label{Leps} \\
	P^\eps(T,x,y,z) 						&=& h(x) , \label{eq:PepsBC}
\end{eqnarray}
where the operator $\L_{(X,Y,Z)}$ is the infinitesimal generator of the  process $(X_t,Y_t,Z_t)$:
\begin{eqnarray*}
	\L_{(X,Y,Z)} &=& r x \frac{\d}{\d x} + \frac{1}{2} f^2(y) z x^2 \frac{\d^2}{\d x^2}
								 + \rho_{xz} \sigma f(y) z x \frac{\d^2}{\d x \d z}  \\
						 & & + \kappa (\theta - z) \frac{\d}{\d z} + \frac{1}{2} \sigma^2 z \frac{\d^2}{\d z^2}  \\
						 & & + \frac{z}{\eps} \left( (m - y) \frac{\d}{\d y} + \nu^2 \frac{\d^2}{\d y^2} \right) \\
						 & & + \frac{z}{\sqrt{\eps}} \left( \rho_{yz} \sigma \nu \sqrt{2} \frac{\d^2}{\d y \d z}
						 		 + \rho_{xy} \nu \sqrt{2} f(y) x \frac{\d^2}{\d x \d y} \right).
\end{eqnarray*}
It will be convenient to separate $\L^\eps$ into groups of like-powers of $1/\sqrt{\eps}$.  To this end, we define the operators $\L_0$, $\L_1$ and $\L_2$ as follows:
\begin{eqnarray}
	\L_0 				&:=& 	  \nu^2 \frac{\d^2}{\d y^2}+ (m - y) \frac{\d}{\d y} \label{eq:L0}, \\
	\L_1 				&:=& 	\rho_{yz} \sigma \nu \sqrt{2}\, \frac{\d^2}{\d y \d z}
										+ \rho_{xy} \nu \sqrt{2}\, f(y) x \frac{\d^2}{\d x \d y}  \label{eq:L1}, \\
	\L_2 				&:=& 	\frac{\d}{\d t}+\frac{1}{2} f^2(y) z x^2 \frac{\d^2}{\d x^2}+  r \left(x \frac{\d}{\d x}-\cdot\right)   \nn \\
     					& & 	+ \frac{1}{2} \sigma^2 z \frac{\d^2}{\d z^2} + \kappa (\theta - z) \frac{\d}{\d z}
					+ \rho_{xz} \sigma f(y) z x \frac{\d^2}{\d x \d z} \label{eq:L2} .
\end{eqnarray}
With these definitions, $\L^\eps$ is expressed as:
\begin{eqnarray}
	\L^\eps &=& 	\frac{z}{\eps} \L_0+ \frac{z}{\sqrt{\eps}} \L_1 + \L_2 . \label{eq:Lexpansion}
\end{eqnarray}
Note that $\L_0$ is the infinitesimal generator of an OU process with unit rate of mean-reversion, and $\L_2$ is the pricing operator of the Heston model with volatility and correlation modulated by $f(y)$.

\section{Asymptotic Analysis} \label{sec:asmptotics}
For a general function $f$, there is no analytic solution to the Cauchy problem (\ref{eq:PepsPDE}--\ref{eq:PepsBC}). Thus, we proceed with an asymptotic analysis as developed in \cite{fouque}.  Specifically, we perform a singular perturbation with respect to the small parameter $\eps$, expanding our solution in powers of $\sqrt{\eps}$
\begin{eqnarray}
	P^\eps &=& P_0 + \sqrt{\eps} P_1 + \eps P_2 + \ldots \,.\label{eq:Pexpansion}
\end{eqnarray}
We now plug (\ref{eq:Pexpansion}) and (\ref{eq:Lexpansion}) into (\ref{eq:PepsPDE}) and (\ref{eq:PepsBC}), and collect terms of equal powers of $\sqrt{\eps}$. 
\subsection*{The Order $1/\eps$ Terms}
Collecting terms of order $1/\eps$ we have the following PDE:
\begin{eqnarray}
	0 &=& z \L_0 P_0 . \label{eq:EpsMinusOnePDE}
\end{eqnarray}
We see from (\ref{eq:L0}) that both terms in $\L_0$ take derivatives with respect to $y$. In fact, $\L_0$ is an infinitesimal generator an consequently zero is an eigenvalue with constant eigenfunctions.   Thus, we seek  $P_0$ of the form
\begin{eqnarray*}
	P_0	&=&	P_0(t,x,z),
\end{eqnarray*}
so that (\ref{eq:EpsMinusOnePDE}) is satisfied.
\subsection*{The Order $1/\sqrt{\eps}$ Terms}
Collecting terms of order $1/\sqrt{\eps}$ leads  to the following PDE
\begin{eqnarray}
	0	&=& z \L_0 P_1 + z \L_1 P_0 \nn \\
		&=& z \L_0 P_1 . \label{eq:EpsMinusOneHalfPDE}
\end{eqnarray}
Note that we have used that $\L_1 P_0=0$, since both terms in $\L_1$ take derivatives with respect to $y$ and $P_0$ is independent of $y$. As above,  we seek $P_1$ of the form
\begin{eqnarray*}
	P_1	&=&	P_1(t,x,z),
\end{eqnarray*}
so that (\ref{eq:EpsMinusOneHalfPDE}) is satisfied.
\subsection*{The Order $1$ Terms}
Matching terms of order $1$ leads  to the following PDE and boundary condition:
\begin{eqnarray}
	0						&=& z \L_0 P_2 + z \L_1 P_1 + \L_2 P_0 \nn \\
		 					&=&	z \L_0 P_2 + \L_2 P_0\label{eq:EpsZeroPDE1} \\
		h(x)				&=&	P_0(T,x,z) . \label{eq:EpsZeroBC}
\end{eqnarray}   
In deriving (\ref{eq:EpsZeroPDE1}) we have used that $\L_1 P_1=0$, since  $\L_1$ takes derivative with respect to $y$ and $P_1$ is independent of $y$.  

Note that (\ref{eq:EpsZeroPDE1}) is a Poisson equation in $y$ with respect to the infinitesimal generator $\L_0$ and with source term $\L_2 P_0$; in solving this equation, $(t,x,z)$  are fixed parameters. In order for this equation to admit solutions with reasonable growth at infinity (polynomial growth), we impose that the source term satisfies the following {\it centering condition}:
\begin{eqnarray}
	0		&=& \< \L_2 P_0 \>
	 		=	\< \L_2 \> P_0 \label{eq:EpsZeroPDE2},
\end{eqnarray}
where we have used the notation
\begin{eqnarray}
	\< g \> &:=& 	\int g(y) \Phi(y) dy \label{eq:centeringcondition} ,
\end{eqnarray}
here $\Phi$ denotes the density 
of the invariant distribution  of the process $Y_t$, which we remind the reader is ${\cal N}(m,\nu^2)$.
Note that in (\ref{eq:EpsZeroPDE2}), we have pulled $P_0(t,x,z)$ out of the linear $\< \cdot \>$ operator since it does not depend on $y$.  

Note that the PDE  (\ref{eq:EpsZeroPDE2}) and the boundary condition  (\ref{eq:EpsZeroBC}) jointly define a Cauchy problem that $P_0(t,x,z)$ must satisfy.  

Using equation (\ref{eq:EpsZeroPDE1}) and the centering condition (\ref{eq:EpsZeroPDE2}) we deduce:
\begin{eqnarray}
	P_2 &=& -\frac{1}{z} \L_0^{-1} \left( \L_2-\< \L_2 \> \right) P_0 , \label{eq:P2}
\end{eqnarray}
where $\L_0^{-1}$ is the inverse operator of $\L_0$ acting on the centered functions.

\subsection*{The Order $\sqrt{\eps}$ Terms}
Collecting terms of order $\sqrt{\eps}$, we obtain the following PDE and boundary condition:
\begin{eqnarray}
	0	&=& z \L_0 P_3 + z \L_1 P_2 + \L_2 P_1	\label{eq:EpsOnePDE1} ,	\\
	0 &=& P_1(T,x,z) \label{eq:EpsOneHalfBC}.
\end{eqnarray}
We note that $P_3(t,x,y,z)$ solves the Poisson equation (\ref{eq:EpsOnePDE1}) in $y$ with respect to $\L_0$.  Thus, we  impose the corresponding centering condition on the source $z \L_1 P_2 + \L_2 P_1$, leading to
  \begin{eqnarray}
	\< \L_2 \> P_1 	&=& 	- \< z \L_1 P_2 \> \label{eq:EpsOnePDE4}	.		
\end{eqnarray}
Plugging  $P_2$, given by (\ref{eq:P2}), into equation (\ref{eq:EpsOnePDE4}) gives:
\begin{eqnarray}
	\< \L_2 \> P_1  &=&  	\A P_0 \label{eq:L2P1AP0}, \\
	\A							&:=& 	\< z \L_1 \frac{1}{z} \L_0^{-1} \left( \L_2 - \< \L_2 \> \right) \>. \label{eq:Adef}
\end{eqnarray}
Note that the PDE  (\ref{eq:L2P1AP0}) and the zero boundary condition  (\ref{eq:EpsOneHalfBC})  define a Cauchy problem that $P_1(t,x,z)$ must satisfy.  
\subsection*{Summary of the Key Results}
We summarize the key results of our asymptotic analysis.  We have written the expansion  (\ref{eq:Pexpansion}) for the solution of the PDE problem (\ref{eq:PepsPDE}--\ref{eq:PepsBC}).
Along the way, he have chosen solutions for $P_0$ and $P_1$ which are of the form
$P_0 = P_0(t,x,z)$ and	$P_1 = P_1(t,x,z)$.
These choices lead us to conclude that $P_0(t,x,z)$ and $P_1(t,x,z)$ must satisfy the following Cauchy problems 
\begin{eqnarray}
	\< \L_2 \> P_0 &=& 0 \label{eq:L2P0=0}, \\
	P_0(T,x,z) &=& h(x) \label{eq:P0=h},
\end{eqnarray}
and
\begin{eqnarray}
	\< \L_2 \> P_1(t,x,z)  &=& \A P_0(t,x,z) \label{eq:L2P1=AP0}, \\
	P_1(T,x,z) &=& 0 \label{eq:P1=0},
\end{eqnarray}
where
\begin{eqnarray}
	\<\L_2\>		&=& 	\frac{\d}{\d t}+ \frac{1}{2} \< f^2 \> z x^2 \frac{\d^2}{\d x^2}  + r \left(x \frac{\d}{\d x}-\cdot\right)   \nn \\
     					& &+ \frac{1}{2} \sigma^2 z \frac{\d^2}{\d z^2} 	+ \kappa (\theta - z) \frac{\d}{\d z} 
     				 	+ \rho_{xz} \sigma \< f \> z x \frac{\d^2}{\d x \d z},\label{eq:L2bracket}
\end{eqnarray}
and $\A$ is given by (\ref{eq:Adef}). Recall that the bracket notation is defined in (\ref{eq:centeringcondition}).

\section{Formulas for $P_0(t,x,z)$ and $P_1(t,x,z)$} \label{sec:P0P1}
In this section we use the results of our asymptotic calculations to find  explicit solutions for $P_0(t,x,z)$ and $P_1(t,x,z)$.

\subsection{Formula for $P_0(t,x,z)$}\label{sec:P0}
Recall that $P_0(t,x,z)$ satisfies a Cauchy problem defined by equations (\ref{eq:L2P0=0}) and (\ref{eq:P0=h}).  

Without loss of generality, we normalize $f$ so that  $\< f^2\> = 1$.  Thus, we rewrite $\<\L_2\>$ given by (\ref{eq:L2bracket})  as follows:  
\begin{eqnarray}
	\<\L_2\>		&=& 	\frac{\d}{\d t}+ \frac{1}{2} z x^2 \frac{\d^2}{\d x^2}  + r \left(x \frac{\d}{\d x}-\cdot\right)   \nn \\
     					& &+ \frac{1}{2} \sigma^2 z \frac{\d^2}{\d z^2} 	+ \kappa (\theta - z) \frac{\d}{\d z} 
     				 	+ \rho \sigma  z x \frac{\d^2}{\d x \d z} \label{eq:<L2>}, \\
					&:=&\L_H,\nn\\
  \rho 				&:=&	\rho_{xz} \< f \>.
\end{eqnarray}
We note that $\rho^2 \leq 1$  since $\<f\>^2 \leq \<f^2\>=1$.  So, $\rho$ can be thought of as an {\it effective correlation} between the Brownian motions in the Heston model obtained in the limit $\eps\to 0$, where $\<\L_2\>=\L_H$, the pricing operator for European options as calculated in the Heston model.
Thus, we see that $P_0(t,x,z)=:P_H(t,x,z)$ is the classical solution for the price of a European option as calculated in the Heston model with effective correlation $\rho=\rho_{xz} \< f \>$.

The derivation of pricing formulas for the Heston model is given in  Appendix \ref{sec:heston}.  Here, we simply state the main result:
\begin{eqnarray}
	P_H(t,x,z)			&=& e^{-r \tau} \frac{1}{2 \pi} \int e^{-i k q} \Ghat(\tau,k,z) \hhat(k) dk, \label{eq:PHsolution} \\
	\tau(t) 				&=& T-t , \label{eq:tau} \\
	q(t,x) 					&=& r(T-t) + \log{x}, \label{eq:q} \\
	\hhat(k)				&=& \int e^{ikq}h(e^q) dq, \label{eq:hhat2} \\
	\Ghat(\tau,k,z)	&=& e^{C(\tau,k) + z D(\tau,k)}, \label{eq:Ghat} \\
	C(\tau,k)				&=& \frac{\kappa \theta}{\sigma^2}
											\left( \left( \kappa  + \rho i k \sigma + d(k) \right) \tau
											- 2 \log \left( \frac{1-g(k) e^{\tau d(k)}}{1-g(k)} \right) \right), \label{eq:C} \\
	D(\tau,k)				&=&	\frac{\kappa + \rho i k \sigma + d(k)}{\sigma^2}
        							\left( \frac{1-e^{\tau d(k)}}{1-g(k)e^{\tau d(k)}} \right), \label{eq:D} \\
	d(k)						&=& \sqrt{\sigma^2 (k^2 - ik) +(\kappa+ \rho i k \sigma )^2} ,\label{eq:d} \\
	g(k)						&=& \frac{\kappa + \rho i k \sigma + d(k)}{\kappa + \rho i k \sigma - d(k)}. \label{eq:g}
\end{eqnarray}
We note that, for certain choices of $h$, the integral in (\ref{eq:hhat2}) may not converge.  For example, a European call with strike $K$ has $h(e^q)=(e^q-K)^+$.  In this case, the integral in (\ref{eq:hhat2}) converges only if we set $k=k_r + i k_i$ where $k_i >1$.  Hence, when evaluating (\ref{eq:PHsolution}, \ref{eq:hhat2}) one must impose $k=k_r + i k_i$, $k_r>1$ and $dk = dk_r$.

\subsection{Formula for $P_1(t,x,z)$}\label{sec:P1}
Recall that $P_1(t,x,z)$ satisfies a Cauchy problem defined by equations (\ref{eq:L2P1=AP0}) and (\ref{eq:P1=0}). 
In order to find a solution for $P_1(t,x,z)$ we must first identify the operator $\A$.  To this end, we introduce two functions, $\phi(y)$ and $\psi(y)$, which solve the following Poisson equations in $y$ with respect to the operator $ \L_0 $:
\begin{eqnarray}
	 \L_0 \phi &=& \frac{1}{2} \left(f^2 - \< f^2 \> \right) \label{eq:phi} ,\\
	 \L_0 \psi &=&   f - \< f \> \label{eq:psi}.
\end{eqnarray}
From equation (\ref{eq:Adef}) we have:
\begin{eqnarray*}
	\A 	&=& \< z \L_1 \frac{1}{z} \L_0^{-1} \left( \L_2 - \< \L_2 \> \right) \>  \\
					&=&	\< z \L_1 \frac{1}{z} \L_0^{-1} \frac{z}{2} \left( f^2 - \< f^2 \> \right) x^2 \frac{\d^2}{\d x^2} \>  \\
					& & + \< z \L_1 \frac{1}{z} \L_0^{-1} \rho_{xz} \sigma z \left( f - \<f\> \right) x \frac{\d^2}{\d x \d z} \>  \\
					&=& z\<  \L_1 \phi(y) x^2 \frac{\d^2}{\d x^2} \>  + \rho_{xz} \sigma z \<  \L_1 \psi(y) x \frac{\d^2}{\d x \d z} \> .
\end{eqnarray*}
Using the definition (\ref{eq:L1}) of $\L_1$, one deduces the following 
 expression for $\A$:
\begin{eqnarray}
	\A  &=& V_1 z x^2 \frac{\d^3}{\d z \d x^2} + V_2 z x \frac{\d^3}{\d z^2 \d x} \nn \\
      & & + V_3 z x \frac{\d}{\d x} \left(x^2 \frac{\d^2}{\d x^2}\right) + V_4 z \frac{\d}{\d z} \left(x \frac{\d}{\d x}\right)^2 ,
      \label{eq:Asimple} \\
	V_1 &=&	\rho_{yz} \sigma \nu \sqrt{2} \< \phi' \>,	\label{eq:V1} \\
	V_2 &=& \rho_{xz} \rho_{yz} \sigma^2 \nu \sqrt{2} \< \psi' \>,	\label{eq:V2} \\
	V_3 &=& \rho_{xy} \nu \sqrt{2} \< f \phi' \>, \label{eq:V3} \\
	V_4 &=&  \rho_{xy} \rho_{xz} \sigma \nu \sqrt{2} \< f \psi' \> .	\label{eq:V4}
\end{eqnarray}
Note that we have introduced four group parameters, $V_i$, $i=1 \ldots 4$, which are constants, and can be obtained by calibrating our model to the market as will be done in Section \ref{sec:data}.

Now that we have expressions for $\A$, $P_H$, and $\L_H$, we are in a position to solve for $P_1(t,x,z)$, which is the solution to the Cauchy problem defined by equations (\ref{eq:L2P1=AP0}) and (\ref{eq:P1=0}).  We leave the details of the calculation to Appendix \ref{sec:P1details}.  Here, we simply present the main result.
\begin{eqnarray}
	P_1(t,x,z) 			&=& \frac{e^{-r \tau} }{2 \pi} \int_\R e^{-ikq} \left( \kappa \theta \fhat_0(\tau,k)+z \fhat_1(\tau,k)\right) \nn \\
									& &\times	\Ghat(\tau,k,z)\hhat(k) dk, \label{eq:P1solution} \\
	\tau(t) 				&=& T-t , \nn \\
	q(t,x) 					&=& r(T-t) + \log{x}, \nn \\
	\hhat(k)				&=& \int e^{ikq}h(e^q) dq, \nn \\
	\Ghat(\tau,k,z)	&=& e^{C(\tau,k) + z D(\tau,k)}, \nn \\
	\fhat_0(\tau,k) &=& \int_0^\tau  \fhat_1(s,k) ds, \label{eq:f0hat} \\
	\fhat_1(\tau,k) &=&	\int_0^\tau b(s,k) e^{A(\tau,k,s)} ds, \label{eq:f1hat} \\
	C(\tau,k)				&=& \frac{\kappa \theta}{\sigma^2}
											\left( \left( \kappa  + \rho i k \sigma + d(k) \right) \tau
											- 2 \log \left( \frac{1-g(k) e^{\tau d(k)}}{1-g(k)} \right) \right) ,\nn \\
	D(\tau,k)				&=&	\frac{\kappa + \rho i k \sigma + d(k)}{\sigma^2}
        							\left( \frac{1-e^{\tau d(k)}}{1-g(k)e^{\tau d(k)}} \right) ,\nn \\
	A(\tau,k,s)			&=& \left(\kappa + \rho \sigma i k + d(k)\right) \frac{1-g(k)}{d(k)g(k)}
											\log \left( \frac{g(k) e^{\tau d(k)}-1}{g(k) e^{s d(k)}-1} \right) \nn \\
									&	&	+ d(k) \left(\tau-s \right) ,\label{eq:A} \\
	d(k)						&=& \sqrt{\sigma^2 (k^2 - ik) +(\kappa+ \rho i k \sigma )^2}\,, \nn \\
	g(k)						&=& \frac{\kappa + \rho i k \sigma + d(k)}{\kappa + \rho i k \sigma - d(k)} ,\nn \\
	b(\tau,k)				&=&  -  \left( V_1 D(\tau,k) \left(-k^2+ ik \right) + V_2 D^2(\tau,k) \left( -ik \right) \right. \nn \\
		        			&	&		\left. + V_3 \left(ik^3 + k^2\right) + V_4 D(\tau,k)\left( -k^2 \right) \right). \label{eq:b}									
\end{eqnarray}
Once again, we note that, depending on the option payoff, evaluating equation (\ref{eq:P1solution}) may require setting $k = k_r + i k_i$ and $dk = dk_r$, as described at the end of subsection \ref{sec:P0}.

\section{Accuracy of the  Approximation} \label{sec:accuracy}

In this section, we prove that the approximation $P^\eps \sim P_0 + \sqrt{\eps} P_1$, where $P_0$ and $P_1$ are defined in the previous sections, is accurate to order $\eps^\alpha$ for any given $\alpha \in (1/2,1)$.  
Specifically, for a European option with a smooth bounded payoff, $h(x)$, and with bounded derivatives, we will show:
\begin{eqnarray}\label{accuracy}
	|P^\eps(t,x,y,z)-\left(P_0(t,x,z)+ \sqrt{\eps}P_1(t,x,z)\right)| &\leq& C\,\eps^\alpha ,
\end{eqnarray}
where $C$ is a  constant which depends on $(y,z)$, but is independent of $\eps$.

We start by defining the remainder term $R^\eps(t,x,y,z)$:
\begin{eqnarray}
	R^\eps = \left(P_0 + \sqrt{\eps} P_1 + \eps P_2 + \eps \sqrt{\eps} P_3 \right) - P^\eps.  \label{eq:remainder}
\end{eqnarray}
Recalling that
\begin{eqnarray*}
	0	&=&	\L^\eps P^\eps, \\
	0	&=&	z \L_0 P_0, \\
	0	&=&	z \L_0 P_1 + z \L_1 P_0 ,\\
	0	&=&	z \L_0 P_2 + z \L_1 P_1 + \L_2 P_0, \\ 
	0	&=&	z \L_0 P_3 + z \L_1 P_2 + \L_2 P_1 ,
\end{eqnarray*}
and applying $\L^\eps$ to $R^\eps$, we obtain that $R^\eps$ must satisfy the following PDE:
\begin{eqnarray}
	\L^\eps R^\eps 
					&=&		\L^{\eps} \left( P_0 + \sqrt{\eps} P_1 + \eps P_2 + \eps \sqrt{\eps} P_3 \right) - \L^\eps P^\eps \nn \\ 
					&=&		\left( \frac{z}{\eps} \L_0 + \frac{z}{\sqrt{\eps}} \L_1 + \L_2 \right)
								\left( P_0 + \sqrt{\eps} P_1 + \eps P_2 + \eps \sqrt{\eps} P_3 \right)  \nn \\
										&=& 	\eps \left( z \L_1 P_3 + \L_2 P_2 +	\sqrt{\eps} \L_2 P_3 \right) \nn	\\
					&=& 	\eps \, F^\eps, \label{eq:RfirstPDE} \\
	F^\eps	&:=&	z \L_1 P_3 + \L_2 P_2	+ \sqrt{\eps} \L_2 P_3, \label{eq:F}
\end{eqnarray}
where we have defined  the $\eps$-dependent source term $F^\eps(t,x,y,z)$.
Recalling that
\begin{eqnarray*}
	P^\eps(T,x,y,z)	&=&	h(x),	\\
	P_0(T,x,z)			&=&	h(x),	\\
	P_1(T,x,z)			&=& 0,								
\end{eqnarray*}
we deduce  from (\ref{eq:remainder}) that
\begin{eqnarray}
	R^\eps(T,x,y,z) &=& \eps P_2(T,x,y,z) + \eps \sqrt{\eps} P_3(T,x,y,z) \nn \\
																		&=& \eps \, G^\eps(x,y,z), \label{eq:RfirstBC} \\
	G^\eps(x,y,z)	&:=& 	P_2(T,x,y,z) + \sqrt{\eps} P_3(T,x,y,z), \label{eq:G}	
\end{eqnarray}
where we have defined the $\eps$-dependent boundary term $G^\eps(x,y,z)$.

Using the expression (\ref{Leps}) for ${\L^\eps}$ we find that $R^\eps(t,x,y,z)$ satisfies the following Cauchy problem with source:
\begin{eqnarray}
	\left( \frac{\d}{\d t} + \L_{X,Y,Z} - r \right) R^\eps &=& \eps {\,} F^\eps, \label{eq:PDEforR} \\
	 R^\eps(T,x,y,z) &=& \eps {\,} G^\eps(x,y,z) .  \label{eq:BCforR}
\end{eqnarray}
Therefore $R^\eps$  admits the following  probabilistic representation:
\begin{eqnarray}
	&&\hskip -1cm R^\eps(t,x,y,z) = 
		\eps\,  \E \bigg[ e^{-r(T-t)} G^\eps(X_T,Y_T,Z_T)  \nn \\
	& &	 - \int_t^T e^{-r(s-t)} F^\eps(s,X_s,Y_s,Z_s) ds \mid X_t=x,Y_t=y,Z_t=z \bigg] .
	\label{eq:StochR}
\end{eqnarray}
In order to bound $R^\eps(T,x,y,z)$, we need  bounds on the growth of $F^\eps(t,x,y,z)$ and $G^\eps(x,y,z)$.  From equation (\ref{eq:G}) we see that $G^\eps(x,y,z)$ contains the functions $P_2(t,x,y,z)$ and $P_3(t,x,y,z)$.  And from  equation (\ref{eq:F}) we see that $F^\eps(t,x,y,z)$ contains terms with the linear operators, $\L_1$ and $\L_2$, acting on $P_2(t,x,y,z)$ and $P_3(t,x,y,z)$.  Thus, to bound $F^\eps(t,x,y,z)$ and $G^\eps(x,y,z)$, we need to obtain growth estimates for $P_2(t,x,y,z)$, $P_3(t,x,y,z)$ and growth estimates for $P_2(t,x,y,z)$ and $P_3(t,x,y,z)$ when linear operators  act upon them.  To do this, we use the following classical result, which can be found in Chapter 5 of \cite{fouque}.
\begin{lemma} \label{th:poisson}
Suppose $\L_0 \chi = g$, $\<g\>=0$ and $|g(y)|<C_1(1+|y|^n)$, then $|\chi(y)|<C_2(1+|y|^n)$ for some $C_2$.  When $n=0$ we have $|\chi(y)|<C_2(1+\log(1+|y|))$.
\end{lemma}
Now, by continuing the asymptotic analysis of Section \ref{sec:asmptotics}, we find that $P_2(t,x,y,z)$ and $P_3(t,x,y,z)$ satisfy Poisson equations in $y$ with respect to the operator, $\L_0$.  We have
\begin{align*}
\L_0 P_2(t,x,y,z)
		&=	\frac{1}{z} \left( -\L_2 + \< \L_2 \> \right) P_0(t,x,z)	,	\\
\L_0 P_3(t,x,y,z)
		&=	\frac{1}{z} \left( -\L_2 + \< \L_2 \> \right) P_1(t,x,z)	+ \left( -\L_1 P_2(t,x,y,z) + \< \L_1 P_2(t,x,y,z) \> \right)	.
\end{align*}
Also note, for any operator, $\M$, of the form
\begin{align}
	\M
		&=	\frac{\d^m}{\d z^m} \prod_{j=1}^N x^{n(j)}\frac{\d^{n(j)}}{\d x^{n(j)} }	, \label{eq:M}
\end{align}
we have $\M \L_0 = \L_0 \M$, because $\L_0$ does not contain $x$ or $z$.  Hence, $\M P_2(t,x,y,z)$ and $\M P_3(t,x,y,z)$ satisfy the following Poisson equations in $y$ with respect to the operator, $\L_0$
\begin{align}
\L_0 \left( \M P_2(t,x,y,z) \right)
		&=	\M \frac{1}{z} \left( -\L_2 + \< \L_2 \> \right) P_0(t,x,z)	,	\label{eq:L0MP2} \\
\L_0 \left( \M P_3(t,x,y,z) \right)
		&=	\M \frac{1}{z} \left( -\L_2 + \< \L_2 \> \right) P_1(t,x,z)	\nn	\\
		&\qquad	+ \M \left( -\L_1 P_2(t,x,y,z) + \< \L_1 P_2(t,x,y,z) \> \right)	.	\nn
\end{align}
Let us   bound functions of the form $\M P_0(t,x,z)$.  Using equations (\ref{eq:PHsolution}) and (\ref{eq:M}), and recalling that $q = r\tau + \log x$ and $\Ghat = e^{C + z D}$, we have
\begin{align*}
\M P_0
	&=	\frac{e^{-r \tau}}{2 \pi} \int \left(\prod_{j=1}^N x^{n(j)}\frac{\d^{n(j)}}{\d x^{n(j)} } e^{-ikq} \right) 
				\left( \frac{\d^m}{\d z^m} e^{C(\tau,k,z)+z D(\tau,k,z)}\right) \hhat(k) dk	\\
	&=	\frac{e^{-r \tau}}{2 \pi} \int e^{-ikq} \left(\prod_{j=1}^N \prod_{l=1}^{n(j)}(-ik-l+1)\right) 
				\left( \left(D(\tau,k,z)\right)^m e^{C(\tau,k,z)+z D(\tau,k,z)}\right) \hhat(k) dk	\\	
	&=	\frac{e^{-r \tau}}{2 \pi} \int \left(\prod_{j=1}^N \prod_{l=1}^{n(j)}(-ik-l+1)\right) 
				 \left(D(\tau,k,z) \right)^m e^{-ikq} \Ghat(\tau,k,z) \hhat(k) dk	.
\end{align*}
We note the following:
\begin{itemize}
	\item By assumption, the option payoff, $h(e^q) \in \cal{S}$, the Schwartz class of rapidly decreasing functions.  It is a fact that the Fourier transform, $\hhat(k) \in \cal{S}$ as well.  This implies that $\left\| k^m \hhat(k) \right\|_\infty < \infty$ for all integers, $m$.  
	\item $\left|\Ghat(\tau,k,z)\right| \leq 1$ for all $\tau \in [0,T]$, $k \in \R$, $z \in \R^+$.  This follows from the fact that $\Ghat(\tau,k,z)$ is the characteristic function, $\E [ \exp (i k Q_T)|X_t = x, Z_t = z]$.
	\item There exists a constant, $C$, such that $|D(\tau,k)| \leq C (1 + |k|)$ for all $\tau \in [0,T]$.
\end{itemize}
It follows that for any $\M$ of the form (\ref{eq:M}) we have the following bound on $\M P_0(t,x,z)$
\begin{align}
|\M P_0(t,x,z)|
	&\leq	\frac{e^{-r \tau}}{2 \pi} \int \left|\prod_{j=1}^N \prod_{l=1}^{n(j)}(-ik-l+1)\right| 
				 \left|D(\tau,k) \right|^m \left|e^{-ikq}\right| \left|\Ghat(\tau,k,z)\right| \left|\hhat(k)\right| dk	\nn \\
	&\leq	\int \left|\prod_{j=1}^N \prod_{l=1}^{n(j)}(-ik-l+1)\right| 
				 \left|D(\tau,k) \right|^m \left|\hhat(k)\right| dk := C < \infty ,	\label{eq:MP0bound}
\end{align}
The constant $C$  depends on $\M$, but is independent of $(t, x, z)$. Using similar techniques, a series of tedious but straightforward calculations leads to the following bounds
\begin{align*}
\left|\M P_1(t,x,z)\right|
	&\leq C (1 + z)	,	\\
\left|\frac{\d}{\d t}\M P_0(t,x,z) \right|
	&\leq	C (1 + z) ,	\\
\left|\frac{\d}{\d t}\M P_1(t,x,z) \right|
	&\leq C (1 + z^2)	,
\end{align*}
where, in each case, $C$ is some finite constant which depends on $\M$, but is independent of $(t,x,z)$.
We are now in a position to bound functions of the form $\M P_2(t,x,y,z)$ and $ \M P_3(t,x,y,z)$.  From equation (\ref{eq:L0MP2}) we have
\begin{align*}
\L_0 \left( \M P_2(t,x,y,z) \right)
		&=	\M \frac{1}{z} \left( -\L_2 + \< \L_2 \> \right) P_0(t,x,z)	\\
		&=	\frac{1}{2} \left( -f^2(y) + \< f^2 \> \right) \M_1 P_0(t,x,z) \\
		&\quad + \rho_{xz} \sigma \left( -f(y) + \< f \> \right) \M_2 P_0(t,x,z)	\\
		&=:	g(t,x,y,z)	,
\end{align*}
where $\M_i$ are of the form (\ref{eq:M}).  Now using the fact that $f(y)$ is bounded and using equation (\ref{eq:MP0bound}) we have
\begin{align*}
\left|g(t,x,y,z)\right|
	\leq C	,
\end{align*}
where $C$ is a constant which is independent of $(t,x,y,z)$.  Hence, using lemma \ref{th:poisson}, there exists a constant, $C$, such that
\begin{align*}
\left|\M P_2(t,x,y,z)\right|
	&\leq	C (1+\log(1+|y|))	.
\end{align*}
Similar, but  more involved calculations, lead to the following bounds:
\begin{align}
\left|\M P_3(t,x,y,z)\right|	,
\left|\frac{\d}{\d t}\M P_2(t,x,y,z)\right|
	&\leq	C (1 + \log(1 + |y|)) (1 + z),	\label{eq:MP3bound}	\\
\left|\frac{\d}{\d y}\M P_2(t,x,y,z)\right|
	&\leq	C ,	\nn	\\
\left|\frac{\d}{\d y}\frac{\d}{\d t}\M P_2(t,x,y,z)\right|	,
\left|\frac{\d}{\d y}\M P_3(t,x,y,z)\right|
	&\leq	C (1 + z),	\nn	\\
\left|\frac{\d}{\d t}\M P_3(t,x,y,z)\right|
	&\leq	C (1 + \log(1 + |y|)) (1 + z^2),	\nn	\\
\left|\frac{\d}{\d y}\frac{\d}{\d t}\M P_3(t,x,y,z)\right|
	&\leq	C (1 + z^2)	.	\label{eq:dydtMP3bound}
\end{align}
  We can now bound $G^\eps(x,y,z)$.  Using equation (\ref{eq:G}) we have
\begin{align}
|G^\eps(x,y,z)|
		&\leq |P_2(T,x,y,z)| + \sqrt{\eps}| P_3(T,x,y,z)| \nn	\\
		&\leq	C_1 (1+\log(1+|y|)) + \sqrt{\eps}	C_2 (1 + \log(1 + |y|)) (1 + z)	\nn	\\
		&\leq	C (1 + \log(1 + |y|)) (1 + z)	.	\label{boundG}
\end{align}
Likewise, using equation (\ref{eq:F}), we have
\begin{align*}
\left|F^\eps(t,x,y,z)\right|
	&\leq	z \left|\L_1 P_3(t,x,y,z)\right| + \left|\L_2 P_2(t,x,y,z)\right|	+ \sqrt{\eps} \left|\L_2 P_3(t,x,y,z)\right|	.
\end{align*}
Each of the above terms can be bounded using equations (\ref{eq:MP3bound}-\ref{eq:dydtMP3bound}).  In particular we find that there exists a constant, $C$, such that
\begin{align}
\left|F^\eps(t,x,y,z)\right|
	&\leq  C(1+\log(1+|y|))(1+z^2)	.	\label{boundF}
\end{align}
Using (\ref{eq:StochR}), the bounds (\ref{boundG}) and (\ref{boundF}), Cauchy-Schwarz inequality, and moments of the $\eps$-independent CIR process $Z_t$ (see for instance \cite{ll}), one obtains:
\begin{align}
	&\left| R^\eps(t,x,y,z)\right|	
	\leq	\eps \, C(z) \, \left(1 + \E_{t,y,z}|Y_T|
	 + \int_t^T \E_{t,y,z}|Y_s|ds \right) ,	\label{eq:RepsBound1}
\end{align}
where $\E_{t,y,z}$ denotes the expectation starting at time $t$ from $Y_t=y$ and $Z_t=z$ under the dynamics (\ref{eq:OU})--(\ref{eq:CIR}). 
Under this dynamics, starting at time zero from $y$, we have
\begin{align}
	Y_t		&=	m + (y-m)e^{-\frac{1}{\eps}\int_0^t Z_sds}
						+\frac{\nu\sqrt{2}}{\sqrt{\eps}} e^{ -\frac{1}{\eps}\int_0^t Z_udu} \int_0^t e^{ \frac{1}{\eps}\int_0^s Z_udu}\nu \sqrt{Z_s}\,dW^y_s	. \label{eq:Yexplicit}
\end{align}
Using the bound established in Appendix \ref{appendixbound}, we have that for any given $\alpha \in (1/2,1)$ there is a constant $C$ such that.
\begin{align}
 \E|Y_t|\leq C\, \eps^{\alpha-1}\,,
 \end{align}
and the error estimate (\ref{accuracy}) follows.

\section*{Numerical Illustration for Call Options}

The result of accuracy above is established for smooth and bounded payoffs. The case of call options, important for implied volatilities and calibration described in the following sections,  would require regularizing the payoff as was done in \cite{proof} in the Black-Scholes case with fast mean-reverting stochastic volatility. Here, in the case of the multi-scale Heston model, we simply provide a numerical illustration of the accuracy of approximation. The full model price is computed by Monte Carlo simulation and the approximated price is given by the formula for the Heston price $P_0$ given in Section \ref{sec:P0}, and our formulas for the correction $\sqrt{\eps}\,P_1$ given in Section \ref{sec:P1}. Note that  the  group parameters $V_i$ needed to compute the correction are calculated from  the parameters of the full model.

in Table \ref{table:MonteCarlo}, we summarize the results of a Monte Carlo simulation for a European call option.  We use a standard Euler scheme, with a time step of $10^{-5}$ years--which is short enough to ensure that $Z_t$ never becomes negative.  For each value of $\eps$ we run $10^5$ sample paths.  The  parameters used in the simulation are: 
\begin{eqnarray*}
& x=100, z=0.24,   r=0.05,  \kappa=1, \theta=1, \sigma=0.39, \rho_{xz}=-0.35,\\
& y=0.06,  m=0.06, \nu=1, 
\rho_{xy}=-0.35,  \rho_{yz}=0.35,\\
& \tau=1, K=100,
\end{eqnarray*}
 and $f(y)=e^{y-m-\nu^2}$ so that $\<f^2\>=1$.
 Note that although $f$ is not bounded, it is a convenient choice because  it allows for analytic calculation of the four group parameters $V_i$ given by (\ref{eq:V1}--\ref{eq:V4}). We only display the value of the largest one, $\sqrt{\eps}\,V_3$, which controls the correction of the skew due to the presence of $\rho_{xy}$. 
 We note that the value of $\sqrt{\eps}\,V_3$ calibrated to data from the S\&P500 in Section \ref{sec:data} is even smaller than those displayed in the Table.


The first line of Table \ref{table:MonteCarlo} corresponds to the case of a pure Heston model ($\eps=0$). Therefore, the value $P_0=21.0831$ is exact (computed with analytic formulas), and it gives us a calibration of the empirical error due to the Monte Carlo simulation ($\widehat{\sigma}_{MC}=0.1166$). Note that this empirical error is consistent across the values of $\eps$ used in the Table.

As expected, the approximated price $P_0+\sqrt{\eps}P_1$, converges, as $\eps\to 0$, to the pure Heston price, and the approximation falls within one standard deviation of the Monte Carlo price for $\eps<10^{-3}$.  This illustrates the accuracy of our approximation for call options.

\begin{table}[ht]
	\centering 
	\begin{tabular}{c c c c c c} 
		$\eps$ 			&	$\sqrt{\eps}\,V_3$ 			&	$P_0+\sqrt{\eps}\,P_1$		&$\widehat{P}_{MC}$			&$\widehat{\sigma}_{MC}$		&$|P_0+\sqrt{\eps}\,P_1-\widehat{P}_{MC}|$	\\
		[1ex] 
		\hline 
		$0$ 				& $0.0000$		& $21.0831$ 	& $21.1591$		&	$0.1166$		&		$0.0760$	\\
		\hline 
		$10^{-4}$ 	& $0.0096$ 		& $21.0055$ 	& $21.0045$ 	&	$0.1153$		&		$0.0010$	\\
		$10^{-3}$ 	& $0.0303$ 		& $20.8546$		& $20.7824$ 	&	$0.1136$		&		$0.0722$	\\
		$10^{-2}$		& $0.0959$ 		& $20.3752$ 	& $18.8250$ 	&	$0.1015$		&		$1.5502$	\\
		$10^{-1}$		& $0.3033$ 		& $18.8538$ 	& $14.8158$ 	&	$0.0866$		&		$4.0380$	\\
		[1ex] 
		\hline 
	\end{tabular}
	\caption{Results of a Monte Carlo simulation for a European call option.  } 
\label{table:MonteCarlo} 
\end{table}

\section{The Multi-Scale Implied Volatility Surface}\label{sec:impliedvolatility}
In this section, we explore how the implied volatility surface produced by our multi-scale model compares to that produced by the Heston model.  To begin, we remind the reader that an approximation to the price of a European option in the multi-scale model can be obtained through the formula
\begin{eqnarray}
	P^\eps 			&\sim& P_0 + \sqrt{\eps}P_1 \nn \\
							&=& 			P_H + P^\eps_1, \nn \\
	P^\eps_1 	&:=& 			\sqrt{\eps}\,P_1 , \nn 
\end{eqnarray}
where we have absorbed the $\sqrt{\epsilon}$ into the definition of $P^\eps_1$ and used $P_0=P_H$, the Heston price. 
Form the formulas for the correction $P_1$, given in Section \ref{sec:P1}, it can be seen that $P_1$ is linear in $V_i$, $i=1,\cdots,4$. Therefore by setting 
\begin{eqnarray}
	V^\eps_i	&=&	\sqrt{\eps}\,V_i	\qquad i=1 \ldots 4 \nn \,,
\end{eqnarray}
the small correction $P^\eps_1$ is given by the same formulas as $P_1$ with the $V_i$ replaced by the $V^\eps_i$.

 It is important to note that, although adding a fast mean-reverting factor of volatility on top of the Heston model introduces five new parameters ($\nu$, $m$, $\eps$, $\rho_{xy}$, $\rho_{yz}$) plus an unknown function $f$ to the dynamics of the stock (see (\ref{eq:Sigma}) and (\ref{eq:OU})), neither knowledge of the values of these five parameters, nor the specific form of the function $f$ is required to price options using our approximation.  The effect of adding a fast mean-reverting factor of volatility on top of the Heston model is \emph{entirely} captured by the four group parameters $V^\eps_i$, which are constants that can be obtained by calibrating the multi-scale model to option prices on the market.      
\par
\begin{figure}[!ht]
	\centering
  	\subfigure[][]{\includegraphics[scale=0.3]{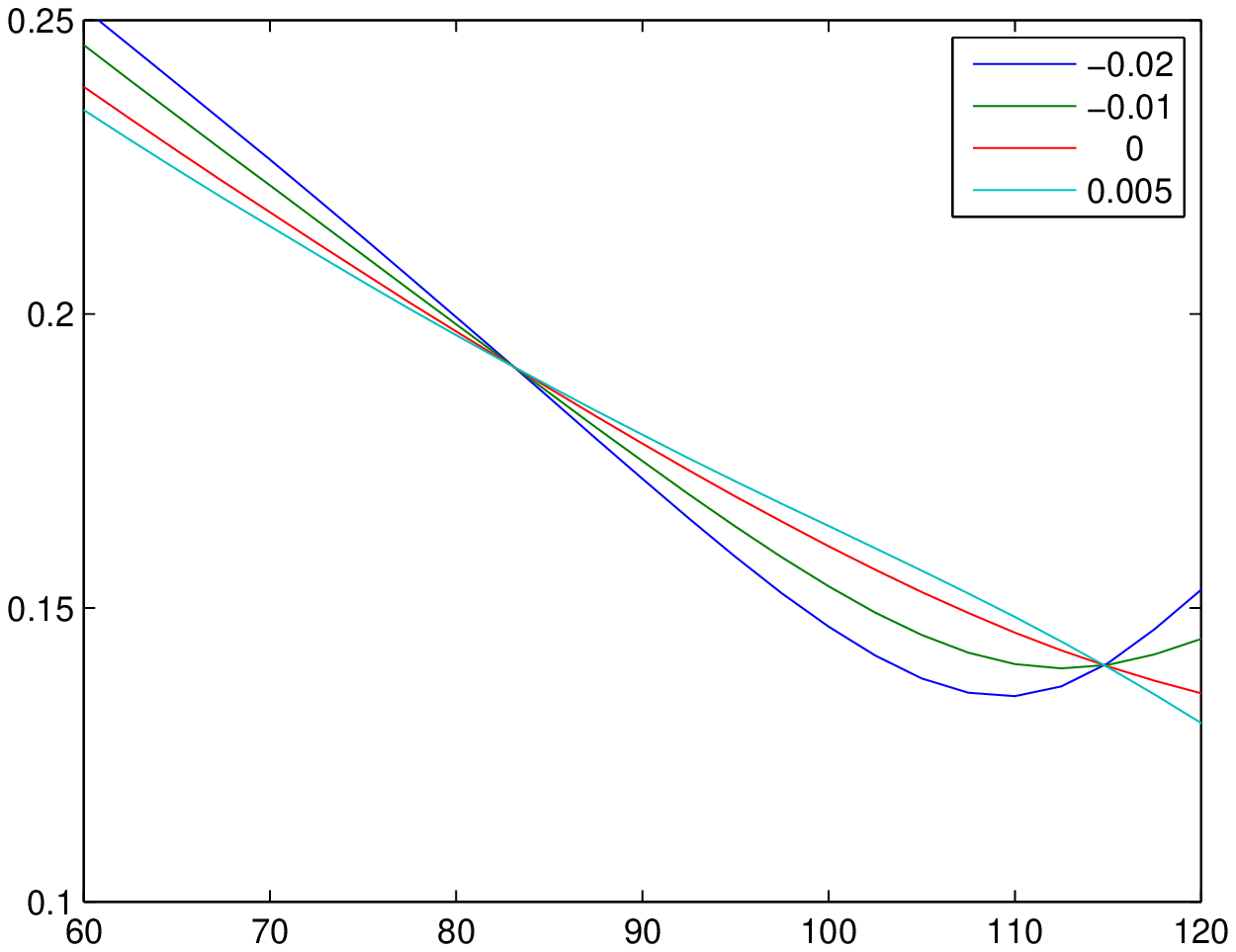} \label{subfig:V1Smile1yrs}}
    \subfigure[][]{\includegraphics[scale=0.3]{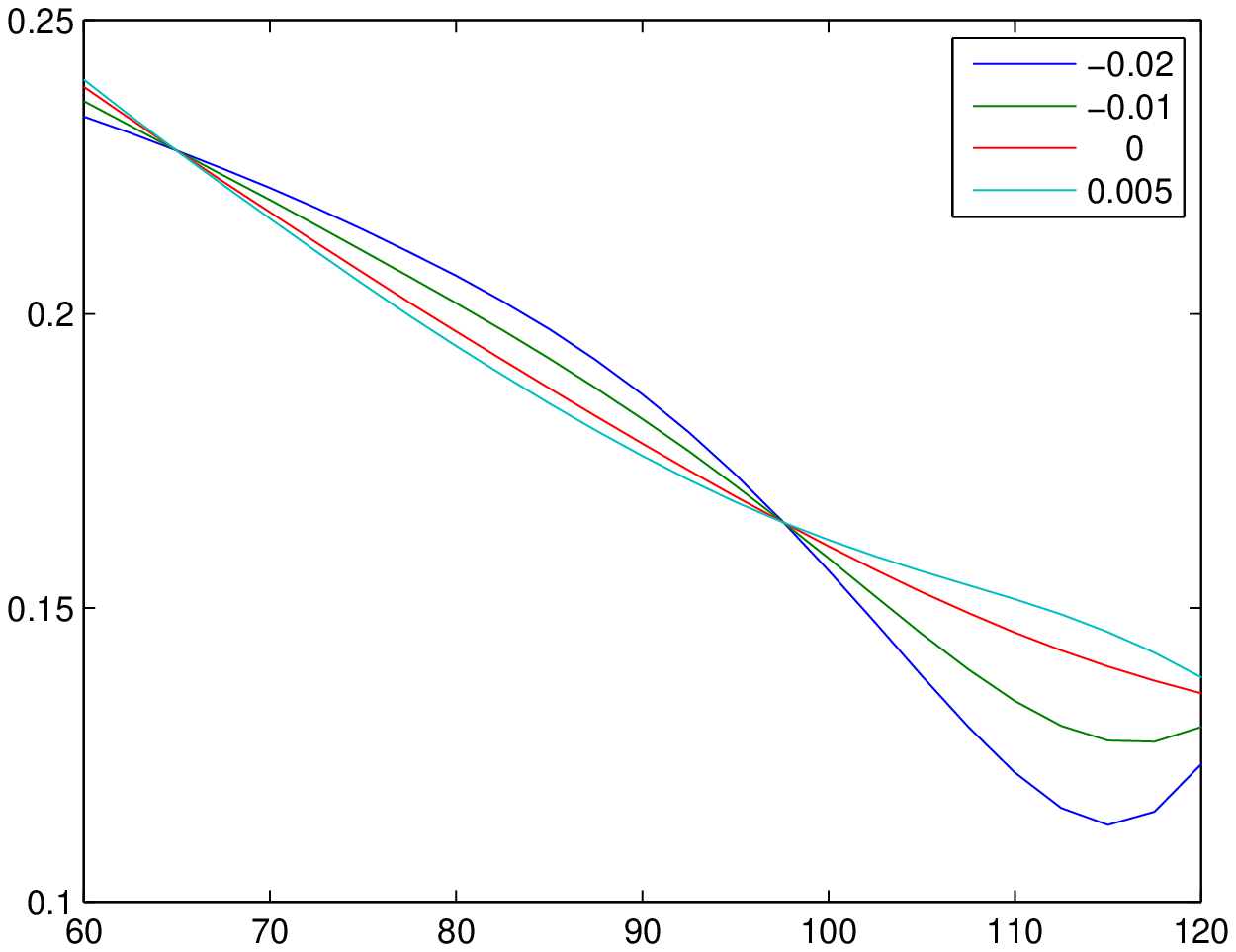} \label{subfig:V2Smile1yrs}}
    \subfigure[][]{\includegraphics[scale=0.3]{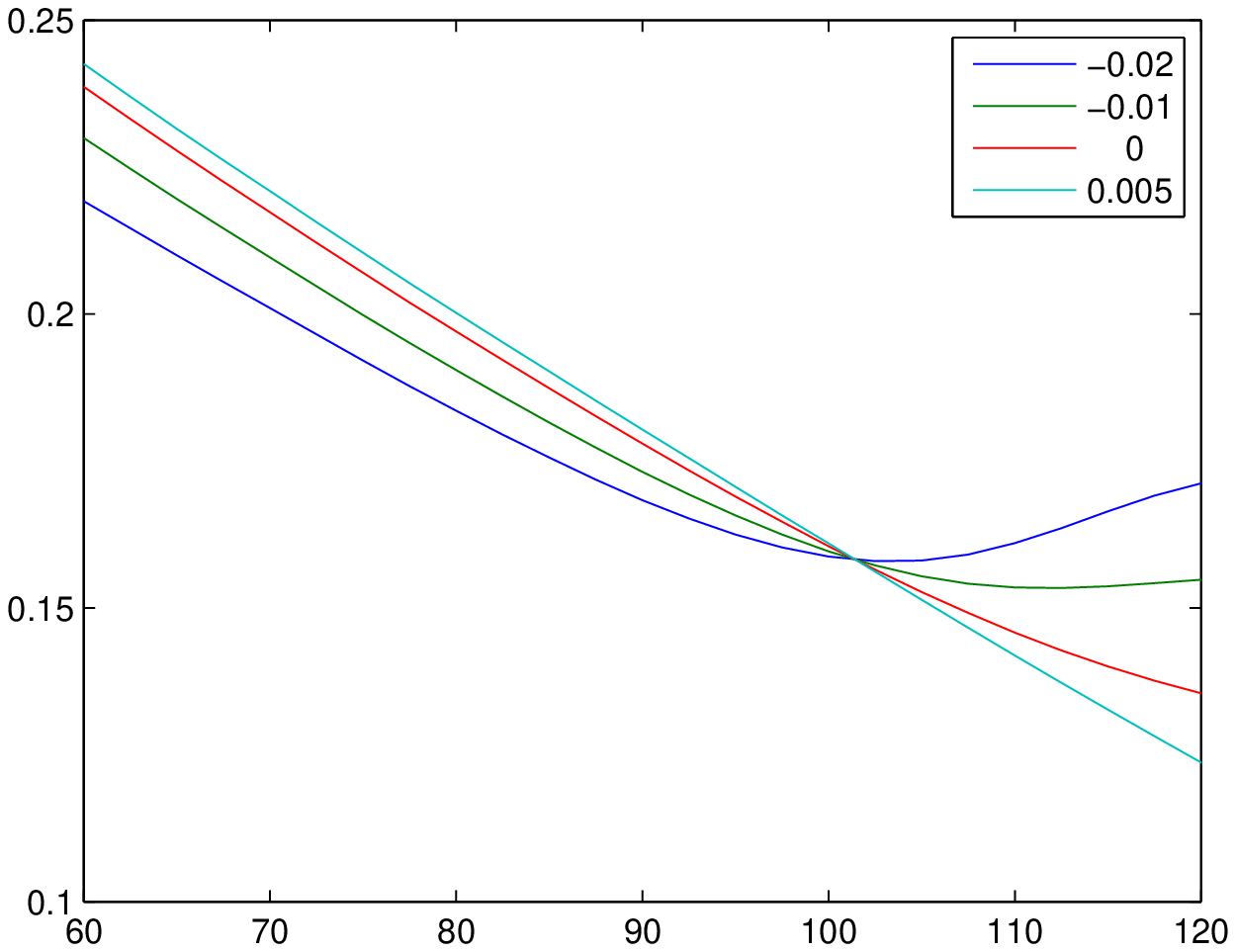} \label{subfig:V3Smile1yrs}}
    \subfigure[][]{\includegraphics[scale=0.3]{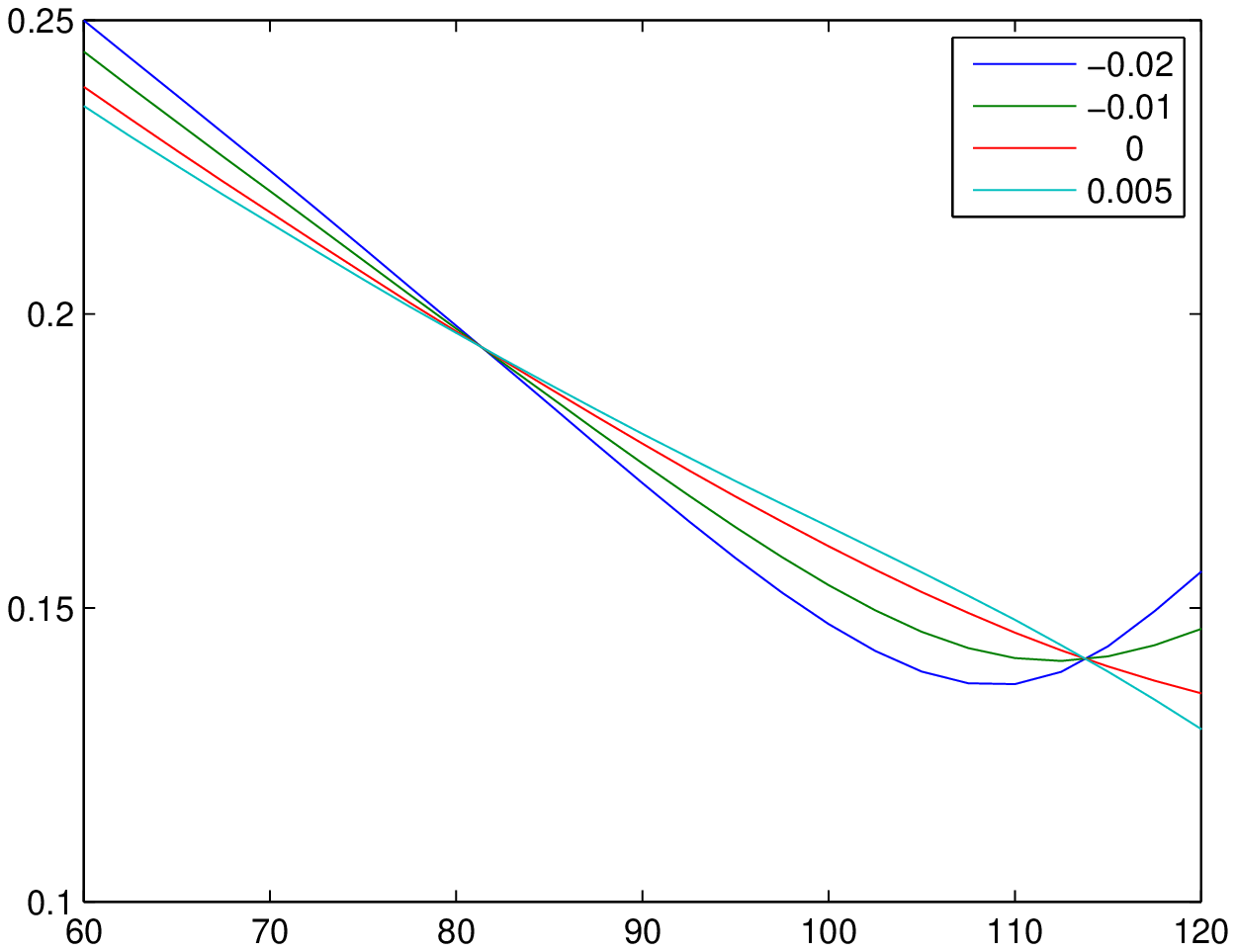} \label{subfig:V4Smile1yrs}}
    \caption[]{Implied volatility curves are plotted as a function of the strike price for European calls in the multi-scale model.  In this example the initial stock price is $x=100$.  The Heston parameters are set to $z=0.04$, $\theta=0.024$, $\kappa=3.4$, $\sigma=0.39$, $\rho_{xz}=-0.64$ and $r=0.0$.  In subfigure \subref{subfig:V1Smile1yrs} we vary only $V^\eps_1$, fixing $V^\eps_i=0$ for $i \neq 1$.  Likewise, in subfigures \subref{subfig:V2Smile1yrs}, \subref{subfig:V3Smile1yrs} and \subref{subfig:V4Smile1yrs}, we vary only $V^\eps_2$, $V^\eps_3$ and $V^\eps_4$ respectively, fixing all other $V^\eps_i=0$.  We remind the reader that, in all four plots, $V^\eps_i=0$ corresponds to the implied volatility curve of the Heston model.}
    \label{fig:Smile1yrs}
\end{figure}

By setting $V^\eps_i=0$ for $i=1,\cdots,4$, we see that $P^\eps_1=0$, $P^\eps = P_H$, and the resulting implied volatility surface, obtained by inverting Black-Scholes formula, corresponds to the implied volatility surface produced by the Heston model.  If we then vary a single $V^\eps_i$ while holding $V^\eps_j=0$ for $j \neq i$, we can see exactly how the multi-scale implied volatility surface changes as a function of each of the $V^\eps_i$.  The results of this procedure are plotted in Figure \ref{fig:Smile1yrs}.

\par
Because they are on the order of $\sqrt{\epsilon}$, typical values of the $V^\eps_i$ are quite small.  However, in order to highlight their effect on the implied volatility surface, the range of values plotted for the $V^\eps_i$ in Figure \ref{fig:Smile1yrs} was intentionally chosen to be large.   It is clear from Figure \ref{fig:Smile1yrs} and from equation (\ref{eq:b}) that each $V^\eps_i$ has a distinct effect on the implied volatility surface.  Thus, the multi-scale model provides considerable flexibility when it comes to calibrating the model to the implied volatility surface produced by options on the market.

\section{Calibration }\label{sec:data}
Denote by $\Theta$ and $\Phi$ the vectors of unobservable parameters in the Heston and Multicale approximation models respectively.
\begin{eqnarray*}
	\Theta 	&=&		(\kappa, \rho, \sigma, \theta, z) ,\\
	\Phi		&=&		(\kappa, \rho, \sigma, \theta, z, V^\eps_1, V^\eps_2, V^\eps_3, V^\eps_4).
\end{eqnarray*}
Let $\sigma(T_i,K_{j(i)})$ be the implied volatility of a call option on the market with maturity date $T_i$ and strike price $K_{j(i)}$.  Note that, for each maturity date, $T_i$, the set of available strikes, $\{K_{j(i)}\}$, varies.  Let $\sigma_H(T_i,K_{j(i)},\Theta)$ be the implied volatility of a call option with maturity date $T_i$ and strike price $K_{j(i)}$ as calculated in the Heston model using parameters $\Theta$.  And let $\sigma_M(T_i,K_{j(i)},\Phi)$ be the implied volatility of call option with maturity date $T_i$ and strike price $K_{j(i)}$ as calculated in the multi-scale approximation using parameters $\Phi$.
\par
We formulate the calibration problem as a constrained, nonlinear, least-squares optimization.  Define the objective functions as
\begin{eqnarray*}
	\Delta_H^2(\Theta)	&=&		\sum_i \sum_{j(i)} \left( \sigma(T_i,K_{j(i)}) - \sigma_H(T_i,K_{j(i)},\Theta) \right)^2,	\\
	\Delta_M^2(\Phi)		&=&		\sum_i \sum_{j(i)} \left( \sigma(T_i,K_{j(i)}) - \sigma_M(T_i,K_{j(i)},\Phi) \right)^2	.
\end{eqnarray*}
We consider $\Theta^{*}$ and $\Phi^{*}$ to be optimal if they satisfy
\begin{eqnarray*}
	\Delta_H^2(\Theta^{*})	&=&	\min_\Theta \Delta_H^2(\Theta)	,\\
	\Delta_M^2(\Phi^{*})	&=&	\min_\Phi \Delta_M^2(\Phi) .
\end{eqnarray*}
It is well-known that that the objective functions, $\Delta_H^2$ and $\Delta_M^2$, may exhibit a number of local minima.  Therefore, if one uses a local gradient method to find $\Theta^{*}$ and $\Phi^{*}$ (as we do in this paper), there is a danger of ending up in a local minima, rather than the global minimum.  Therefore, it becomes important to make a good initial guess for $\Theta$ and $\Phi$, which can be done by visually tuning the Heston parameters to match the implied volatility surface and setting each of the $V_i^\eps=0$.  In this paper, we calibrate the Heston model first to find $\Theta^{*}$.  Then, for the multi-scale model we make an initial guess $\Phi=(\Theta^{*},0,0,0,0)$ (i.e. we set the $V^\eps_i = 0$ and use $\Theta^{*}$ for the rest of the parameters of $\Phi$).  This is a logical calibration procedure because the $V^\eps_i$, being of order $\sqrt{\eps}$, are intended to be small parameters.
\par
\begin{figure}[!ht]
	\centering
  	\includegraphics[scale=0.7]{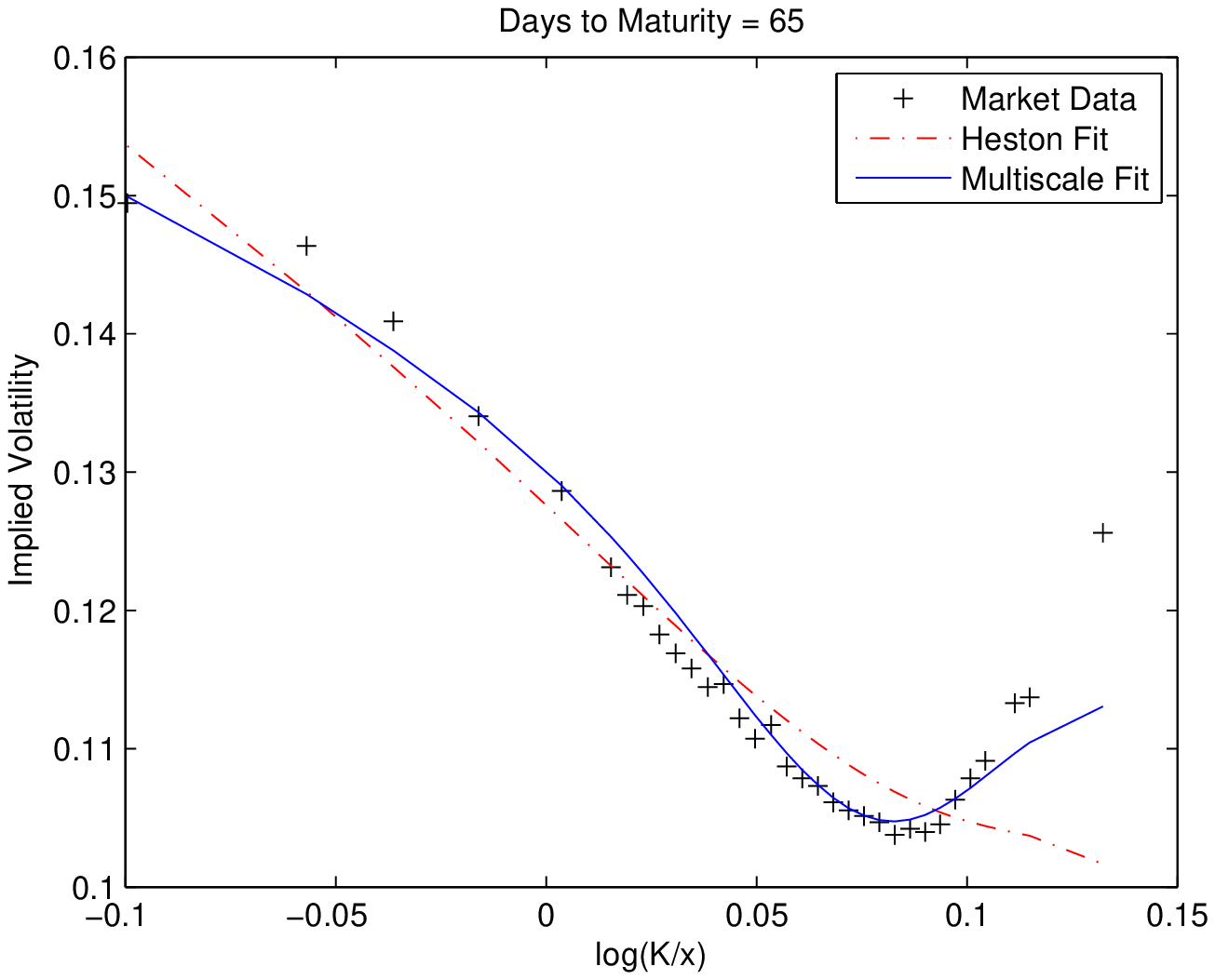}
    \caption[]{SPX Implied Volatilities from May 17, 2006}
    \label{fig:DTM=65}
\end{figure}
\begin{figure}[!ht]
	\centering
  	\includegraphics[scale=0.7]{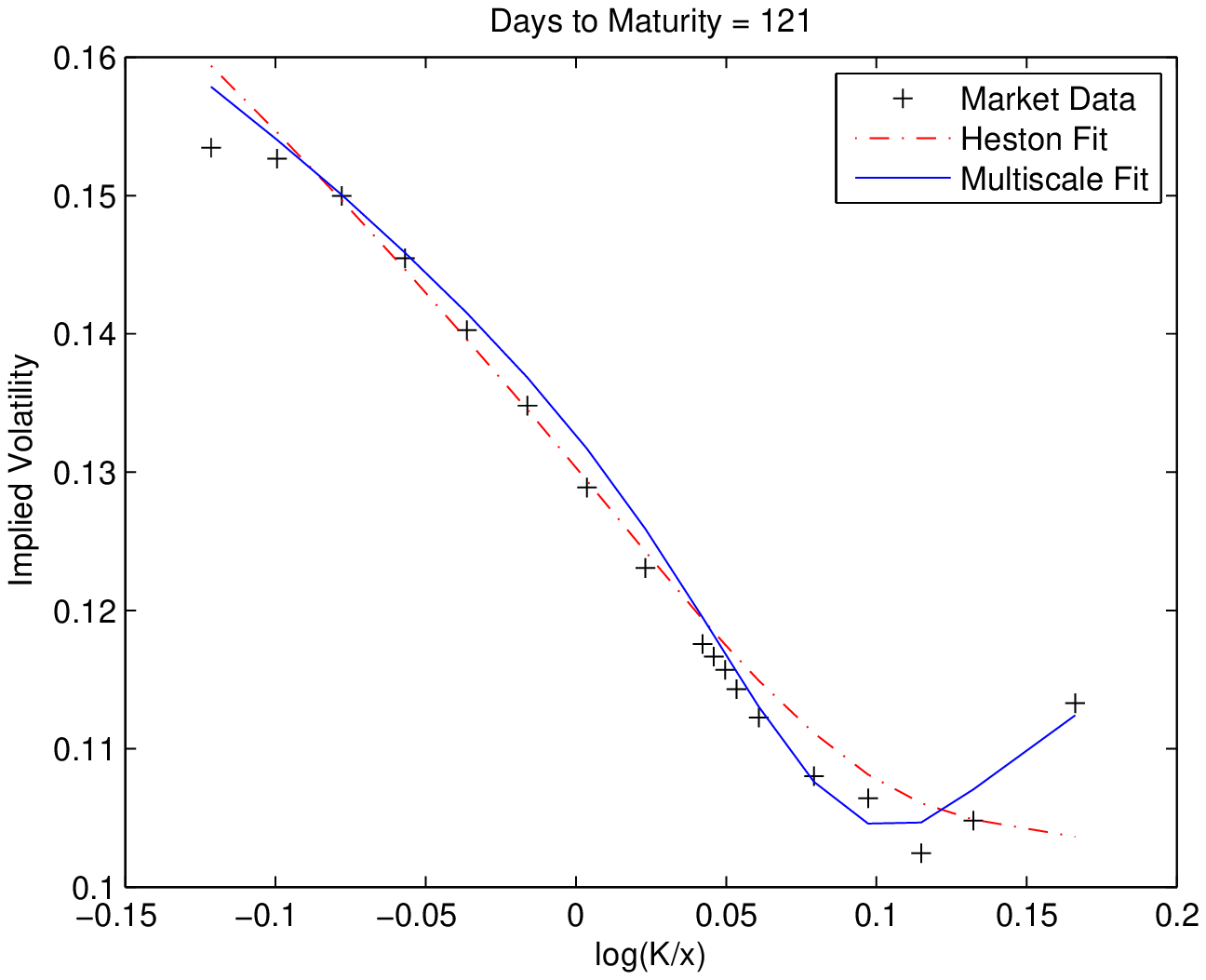}
    \caption[]{SPX Implied Volatilities from May 17, 2006}
		\label{fig:DTM=121}
\end{figure}
\begin{figure}[!ht]
	\centering
  	\includegraphics[scale=0.7]{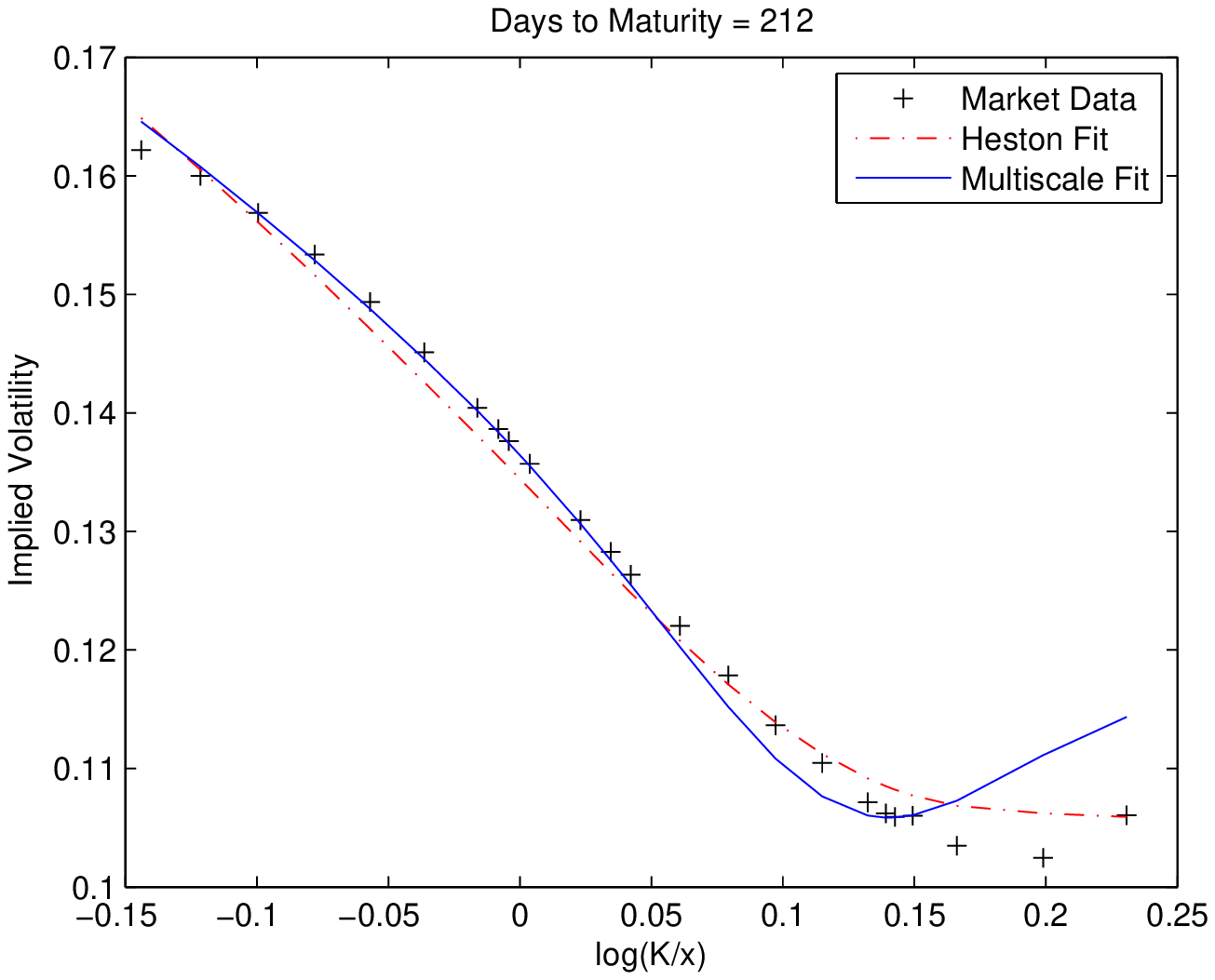}
    \caption[]{SPX Implied Volatilities from May 17, 2006}
    \label{fig:DTM=212}
\end{figure}
\begin{figure}[!ht]
	\centering
  	\includegraphics[scale=0.7]{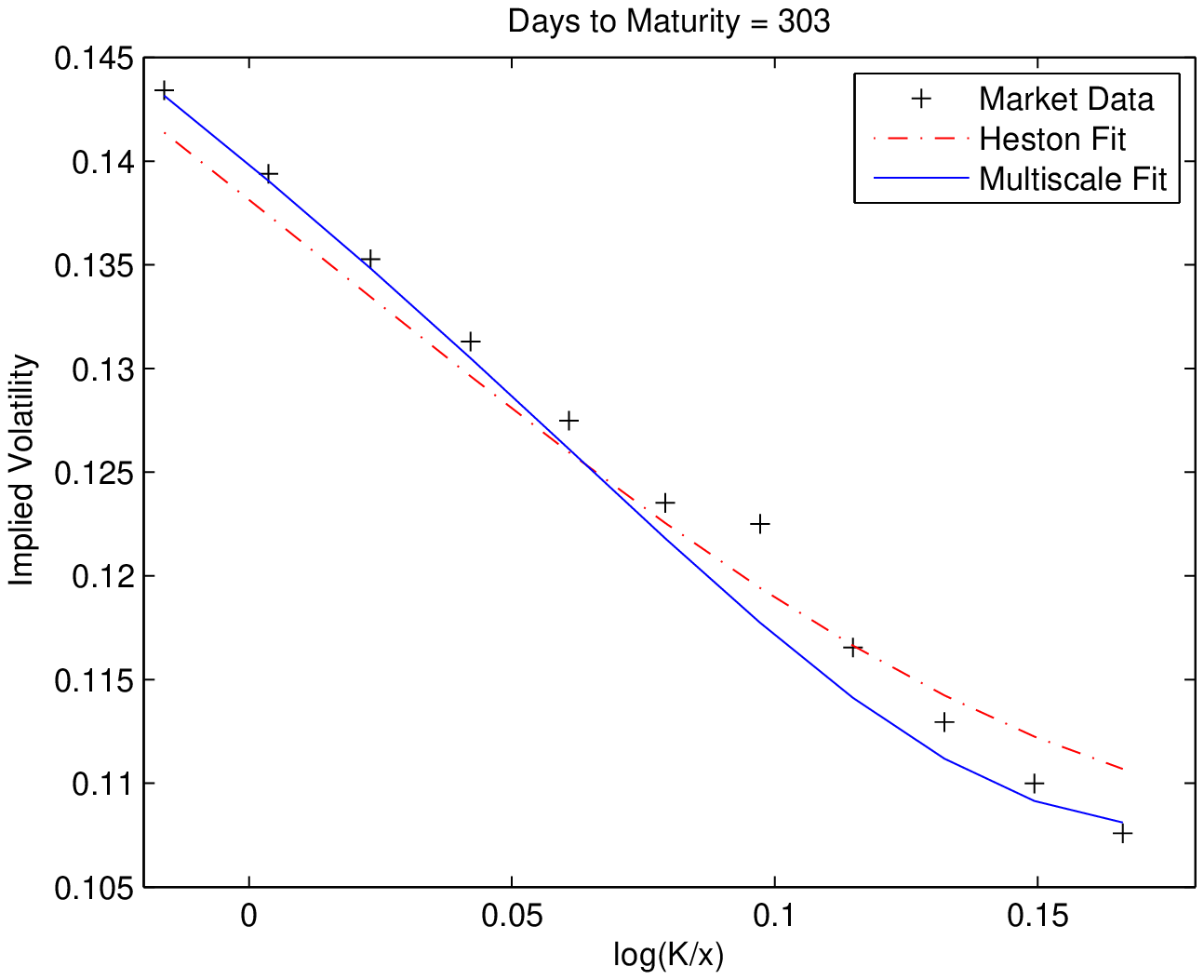}
    \caption[]{SPX Implied Volatilities from May 17, 2006}
    \label{fig:DTM=303}
\end{figure}
\begin{figure}[!ht]
	\centering
  	\includegraphics[scale=0.7]{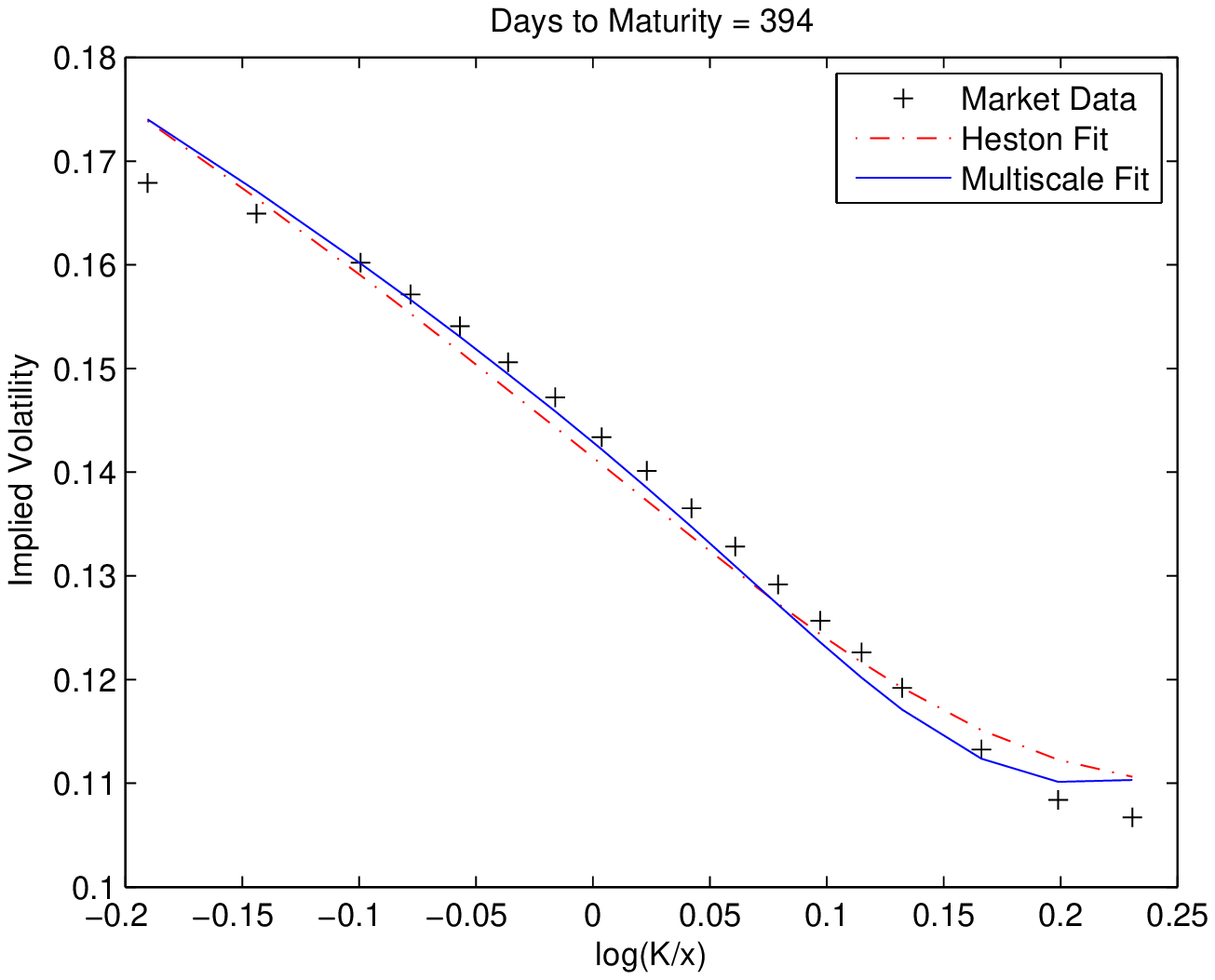}
    \caption[]{SPX Implied Volatilities from May 17, 2006}
		\label{fig:DTM=394}
\end{figure}
\begin{figure}[!ht]
	\centering
  	\includegraphics[scale=0.7]{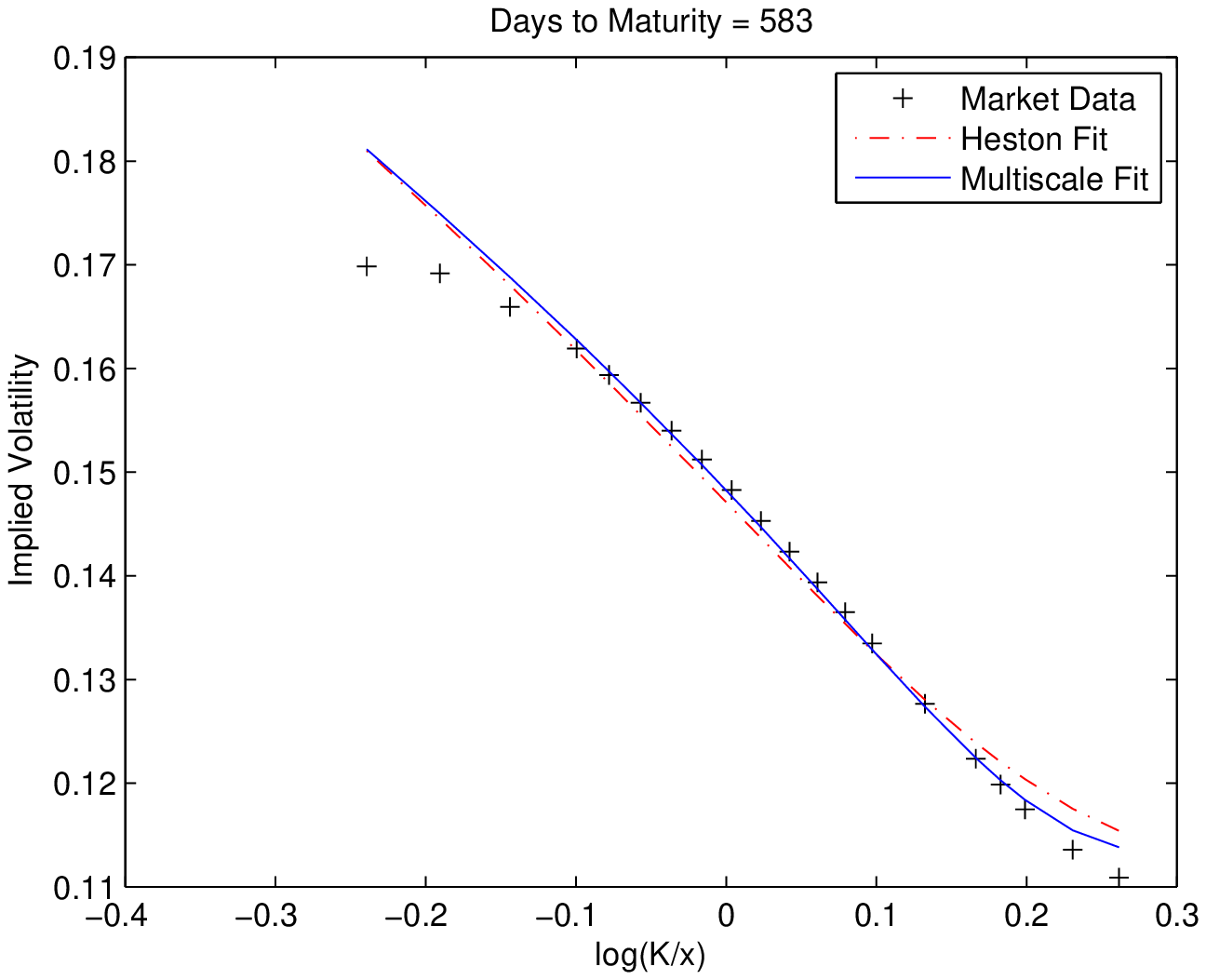}
    \caption[]{SPX Implied Volatilities from May 17, 2006}
		\label{fig:DTM=583}
\end{figure}
\begin{figure}[!ht]
	\centering
  	\includegraphics[scale=0.7]{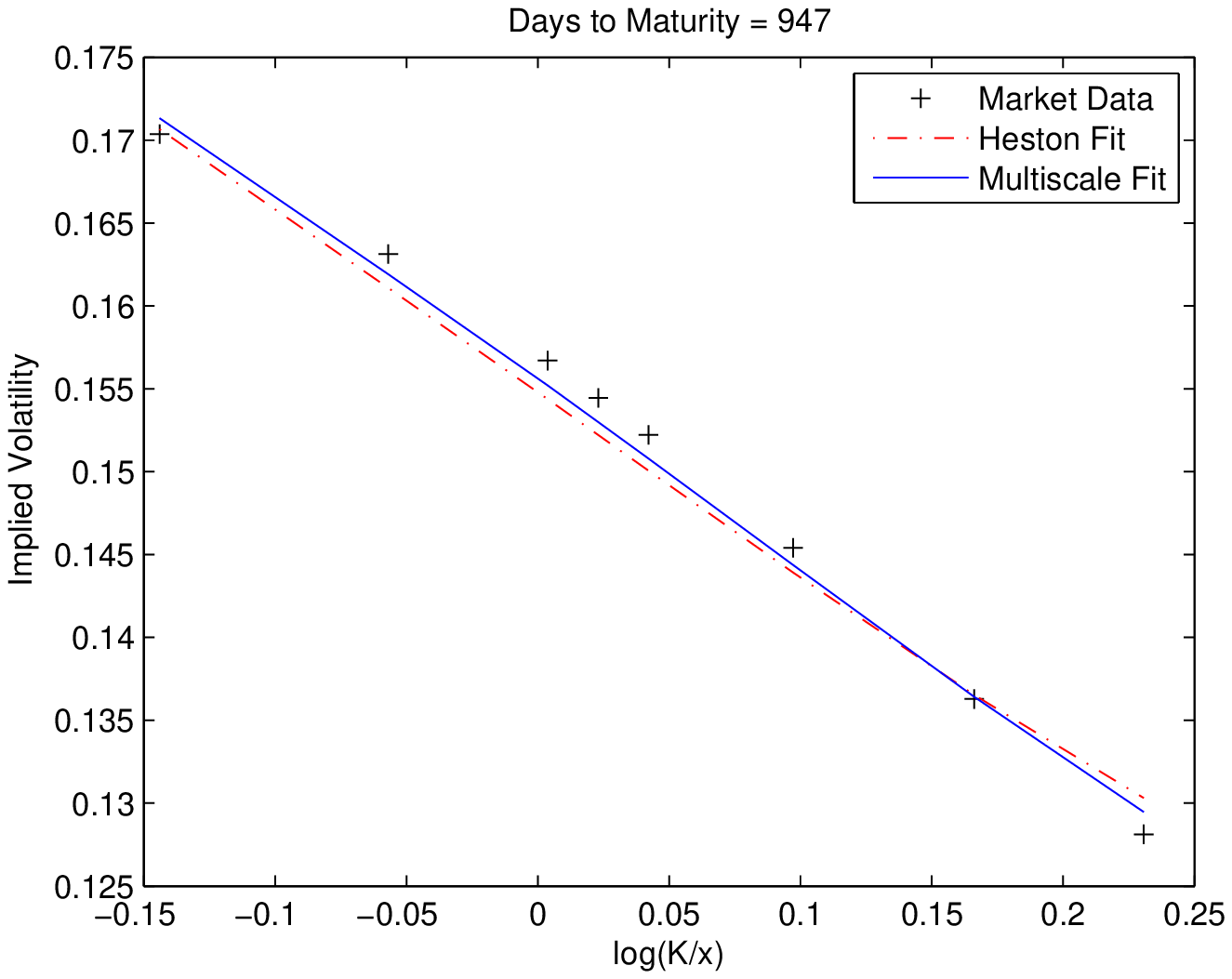}
    \caption[]{SPX Implied Volatilities from May 17, 2006}
		\label{fig:DTM=947}
\end{figure}

The data we consider consists of call options on the S\&P500 index (SPX) taken from May 17, 2006.  We limit our data set to options with maturities greater than $45$ days, and with open interest greater than $100$.  We use the yield on the nominal 3-month, constant maturity, U.S. Government treasury bill as the risk-free interest rate.  And we use a dividend yield on the S\&P 500 index taken directly from the Standard \& Poor's website (www.standardandpoors.com).  In Figures \ref{fig:DTM=65} through \ref{fig:DTM=947}, we plot the implied volatilities of call options on the market, as well as the calibrated implied volatility curves for the Heston and multi-scale models.  We would like to emphasize that, although the plots are presented maturity by maturity, they are the result of a single calibration procedure that uses the entire data set.
\par
From Figures \ref{fig:DTM=65} through \ref{fig:DTM=947}, it is apparent to the naked eye that the multi-scale model represents a vast improvement over the Heston model--especially, for call options with the shortest maturities.  In order to quantify this result we define marginal residual sum of squares
\begin{eqnarray*}
	\bar{\Delta}_H^2(T_i)		&=&		\frac{1}{N(T_i)} \sum_{j(i)} \left( \sigma(T_i,K_{j(i)}) - \sigma_H(T_i,K_{j(i)},\Theta^{*}) \right)^2	,\\
	\bar{\Delta}_M^2(T_i)		&=&		\frac{1}{N(T_i)}	\sum_{j(i)} \left( \sigma(T_i,K_{j(i)}) - \sigma_M(T_i,K_{j(i)},\Phi^{*}) \right)^2 ,\\
\end{eqnarray*}
where $N(T_i)$ is the number of different calls in the data set that expire at time $T_i$ (i.e. $N(T_i) = \#\{K_{j(i)}\}$).  A comparison of $\bar{\Delta}_H^2(T_i)$ and $\bar{\Delta}_M^2(T_i)$ is given in Table \ref{table:statistics}.  The table confirms what is apparent to the naked eye--namely, that the multi-scale model fits the market data significantly better than the Heston model for the two shortest maturities, as well as the longest maturity.
\begin{table}[ht]
	\centering 
	\begin{tabular}{c c c c} 
		$T_i-t$ (days) & $\bar{\Delta}_H^2(T_i)$ & $\bar{\Delta}_M^2(T_i)$ & $\bar{\Delta}_H^2(T_i)/\bar{\Delta}_M^2(T_i)$ \\
		[1ex] 
		\hline 
		$65$ 		& $29.3 \times 10^{-6}$			& $7.91 \times 10^{-6}$ 		& $3.71$		\\
		$121$ 	& $10.2 \times 10^{-6}$ 		& $3.72 \times 10^{-6}$ 		& $2.73$ 		\\
		$212$ 	& $4.06 \times 10^{-6}$ 		& $8.11 \times 10^{-6}$			& $0.51$ 		\\
		$303$		& $3.93 \times 10^{-6}$ 		& $3.51 \times 10^{-6}$ 		& $1.12$ 		\\
		$394$		& $7.34 \times 10^{-6}$ 		& $5.17 \times 10^{-6}$ 		& $1.42$ 		\\
		$583$		& $11.3 \times 10^{-6}$ 		& $9.28 \times 10^{-6}$ 		& $1.22$ 		\\
		$947$		& $3.31 \times 10^{-6}$ 		& $1.47 \times 10^{-6}$ 		& $2.25$ 		\\
		[1ex] 
		\hline 
	\end{tabular}
	\caption{Residual sum of squares for the Heston and the Multi-Scale models at several maturities.} 
\label{table:statistics} 
\end{table}


\appendix
\section{Heston Stochastic Volatility Model} \label{sec:heston}
There are a number of excellent resources where one can read about the Heston stochastic volatility model---so many, in fact, that a detailed review of the model would seem superfluous.  However, in order to establish some notation, we will briefly review the dynamics of the Heston model here, as well as show our preferred method for solving the corresponding European option pricing problem.  The notes from this section closely follow \cite{shaw}.  The reader should be aware that a number of the equations developed in this section are referred to throughout the main text of this paper.
\par
Let $X_t$ be the price of a stock.  And denote by $r$ the risk-free rate of interest.  Then, under the risk-neutral probability measure, $\P$, the Heston model takes the following form:
\begin{eqnarray*}
	dX_t &=& r X_t dt + \sqrt{Z_t}\, X_t dW_t^x\,, \\
	dZ_t &=& \kappa \left( \theta - Z_t \right) dt + \sigma \sqrt{Z_t}\, dW_t^z\,, \\
	d\<W^x,W^z\>_t &=& \rho dt .
\end{eqnarray*}
Here, $W_t^x$ and $W_t^z$ are one-dimensional Brownian motions with correlation $\rho$, such that $|\rho| \leq 1$.  The process, $Z_t$, is the stochastic variance of the stock. And, $\kappa$, $\theta$ and $\sigma$ are positive constants satisfying $2 \kappa \theta \geq \sigma^2$; assuming $Z_0>0$, this ensures that $Z_t$ remains positive for all $t$.
\par
We denote by $P_H$ the price of a European option, as calculated under the Heston framework.  As we are already under the risk-neutral measure, we can express $P_H$ as an expectation of the option payoff, $h(X_T)$, discounted at the risk-free rate.
\begin{eqnarray*}
	P_H(t,x,z) &=& \E \left[ \left. e^{-r(T-t)} h(X_T) \right| X_t=x, Z_t=z \right].
\end{eqnarray*} 
Using the Feynman-Kac formula, we find that $P_H(t,x,z)$ must satisfy the following PDE and boundary condition:
\begin{eqnarray}
	\L_H P_H(t,x,z) &=& 	0,  \label{eq:PDEforPH} \\
	P_H(T,x,z) 			&=& 	h(x), \label{eq:BCforPH} \\
	\L_H					 	&=&  \frac{\d}{\d t} - r + r x \frac{\d}{\d x} + \frac{1}{2} z x^2 \frac{\d^2}{\d x^2} \nn \\
   		  					& & 	+ \kappa \left( \theta - z \right) \frac{\d}{\d z} + \frac{1}{2} \sigma^2 z \frac{\d^2}{\d z^2} \nn \\
     							& & 	+ \rho \sigma z x \frac{\d^2}{\d x \d z} \label{eq:LH}\,.
\end{eqnarray}
In order to find a solution for $P_H(t,x,z)$, it will be convenient to transform variables as follows:
\begin{eqnarray*}
	\tau(t) &=& T-t ,\\
	q(t,x) &=& r(T-t) + \log{x}, \\
	P_H(t,x,z) &=& P_H'(\tau(t),q(t,x),z) e^{-r\tau(t)} .
\end{eqnarray*}
This transformation leads us to the following PDE and boundary condition for $P_H'(\tau,q,z)$:
\begin{eqnarray}
	\L_H' P_H'(\tau, q, z)	&=& 0, \nn \\
	\L_H' 	&=& -\frac{\d}{\d \tau} + \frac{1}{2} z \left(\frac{\d^2}{\d q^2} - \frac{\d}{\d q} \right) 
					+ \rho \sigma z \frac{\d^2}{\d q \d z} \nn \\
					& &	 + \frac{1}{2} \sigma^2 z \frac{\d^2}{\d z^2}
					     + \kappa \left( \theta - z \right)\frac{\d}{\d z}, \label{eq:LHprime} \\
	P_H'(0,q,z) &=& h(e^q) . \nn
\end{eqnarray}
We will find a solution for $P_H'$ through the method of Green's functions.  Denote by $\delta(q)$ the Dirac delta function, and let $G(\tau,q,z)$, the Green's function, be the solution to the following Cauchy problem:
\begin{eqnarray}
	\L_H'G(\tau,q,z) 	&=& 	0 	,				\label{eq:GPDE}\\
	G(0,q,z) 					&=& 	\delta(q). 	\label{eq:GBC}
\end{eqnarray}
Then,
\begin{eqnarray*}
	P_H'(\tau,q,z) &=& \int_\R G(\tau,q-p,z) h(e^p) dp .
\end{eqnarray*}
Now, let $\Phat_H(\tau,k,z)$, $\Ghat(\tau,k,z)$ and $\hhat(k)$ be the Fourier transforms of $P_H'(\tau,q,z)$ $G(\tau,q,z)$ and $h(e^q)$ respectively.
\begin{eqnarray*}
	\Phat_H(\tau,k,z) &=& \int_\R e^{i k q} P_H'(\tau,q,z) dq ,\\
	\Ghat(\tau,k,z) &=& \int_\R e^{i k q} G(\tau,q,z) dq ,\\
	\hhat(k) &=& \int_\R e^{i k q} h(e^q) dq.
\end{eqnarray*}
Then, using the convolution property of Fourier transforms we have:
\begin{eqnarray*}
	P_H'(\tau,q,z) &=& \frac{1}{2 \pi} \int_\R e^{-ikq} \Phat_H(\tau,k,z) dk	\\
								 &=& \frac{1}{2 \pi} \int_\R e^{-ikq} \Ghat(\tau,k,z) \hhat(k) dk .
\end{eqnarray*}
Multiplying equations (\ref{eq:GPDE}) and (\ref{eq:GBC}) by $e^{ikq'}$ and integrating  over $\R$ in $q'$, we find that $\Ghat(\tau,k,z)$ satisfies the following Cauchy problem:
\begin{eqnarray}
	\Lhat_H \Ghat(\tau,k,z) &=& 0, \label{eq:GhatPDE} \\
	\Lhat_H &=& 	-\frac{\d}{\d \tau} + \frac{1}{2} z \left(-k^2 + i k \right) + \frac{1}{2} \sigma^2 z \frac{\d^2}{\d z^2} \nn \\
					& & 	+ \left( \kappa \theta - \left(\kappa + \rho \sigma i k \right) z  \right) \frac{\d}{\d z}, \nn \\ 
	\Ghat(0,k,z) &=& 1 . \label{eq:GhatBC}
\end{eqnarray}
Now, an ansatz: suppose $\Ghat(\tau,k,z)$ can be written as follows:
\begin{eqnarray}
	\Ghat(\tau,k,z) = e^{C(\tau,k)+z D(\tau,k)} . \label{eq:GhatAnsatz}
\end{eqnarray}
Substituting (\ref{eq:GhatAnsatz}) into (\ref{eq:GhatPDE}) and (\ref{eq:GhatBC}), and collecting terms of like-powers of $z$, we find that $C(\tau,k)$ and $D(\tau,k)$ must satisfy the following ODE's
\begin{eqnarray}
	\frac{d C}{d \tau}(\tau,k) &=& \kappa \theta D(\tau,k), \label{eq:ODEforC} \\
	C(0,k) &=& 0 ,\label{eq:BCforC} \\
	\frac{d D}{d \tau}(\tau,k) &=& \frac{1}{2} \sigma^2 D^2(\tau,k) - \left( \kappa + \rho \sigma i k \right) D(\tau,k)
                       + \frac{1}{2} \left( -k^2 + ik \right), \label{eq:ODEforD} \\
  D(0,k) &=& 0 . \label{eq:BCforD}     
\end{eqnarray}
Equations (\ref{eq:ODEforC}), (\ref{eq:BCforC}), (\ref{eq:ODEforD}) and (\ref{eq:BCforD}) can be solved analytically.  Their solutions, as well as the final solution to the European option pricing problem in the Heston framework, are given in (\ref{eq:PHsolution}--\ref{eq:g}).

\section{Detailed solution for $P_1(t,x,z)$} \label{sec:P1details}
In this section, we show how to solve for $P_1(t,x,z)$, which is the solution to the Cauchy problem defined by equations (\ref{eq:L2P1=AP0}) and (\ref{eq:P1=0}).  For convenience, we repeat these equations here with the notation $\L_H=\<\L_2\>$ and $P_H=P_0$:
\begin{eqnarray}
	\L_H P_1(t,x,z)  &=& \A P_H(t,x,z) , \label{eq:P1PDEreminder} \\
	P_1(T,x,z) &=& 0.										\label{eq:P1BCreminder}
\end{eqnarray}
We remind the reader that $\A$ is given by equation (\ref{eq:Asimple}), $\L_H$ is given by equation (\ref{eq:<L2>}), and $P_H(t,x,z)$ is given by equation (\ref{eq:PHsolution}).
It will be convenient in our analysis to make the following variable transformation:
\begin{eqnarray}
	P_1(t,x,z) &=& P_1'(\tau(t),q(t,x),z) e^{-r\tau}, \label{eq:P1andP1prime} \\
	\tau(t) 				&=& T-t , \nn \\
	q(t,x) 					&=& r(T-t) + \log{x}, \nn
\end{eqnarray}
We now substitute equations (\ref{eq:PHsolution}), (\ref{eq:Asimple}) and (\ref{eq:P1andP1prime}) into equations (\ref{eq:P1PDEreminder}) and (\ref{eq:P1BCreminder}), which leads us to the following PDE and boundary condition for $P_1'(\tau,q,z)$:
\begin{eqnarray}
	\L_H' P_1'(\tau,q,z) 
		&=& \A' \frac{1}{2 \pi} \int e^{-ikq} \Ghat(\tau,k,z) \hhat(k) dk \label{eq:P1primePDE}, \\
	\L_H'
	 	&=& -\frac{\d}{\d \tau} + \frac{1}{2} z \left(\frac{\d^2}{\d q^2} - \frac{\d}{\d q} \right) 
				+ \rho \sigma z \frac{\d^2}{\d q \d z} \nn \\
		& &	+ \frac{1}{2} \sigma^2 z \frac{\d^2}{\d z^2}
	    	+ \kappa \left( \theta - z \right) ,\nn \\
	\A'
		&=& V_1 z \frac{\d}{\d z} \left(\frac{\d^2}{\d q^2}-\frac{\d}{\d q} \right) + V_2 z \frac{\d^3}{\d z^2 \d q} \nn \\
    & & + V_3 z \left(\frac{\d^3}{\d q^3}-\frac{\d^2}{\d q^2} \right) + V_4 z \frac{\d^3}{\d z \d q^2}, \nn \\
  P_1'(0,q,z)
  	&=& 0. \label{eq:P1primeBC}
\end{eqnarray}
Now, let $\Phat_1(\tau,k,z)$ be the Fourier transform of $P_1'(\tau,q,z)$
\begin{eqnarray*}
	\Phat_1(\tau,k,z) &=& \int_\R e^{i k q} P_1'(\tau,q,z) dq.
\end{eqnarray*}
Then, 
\begin{eqnarray}
	P_1'(\tau,q,z) &=& \frac{1}{2 \pi} \int_\R e^{-ikq} \Phat_1(\tau,k,z) dk .\label{eq:P1primeFourierRep}
\end{eqnarray}
Multiplying equations (\ref{eq:P1primePDE}) and (\ref{eq:P1primeBC}) by $e^{ikq'}$ and integrating in $q'$ over $\R$, we find that $\Phat_1(\tau,k,z)$ satisfies the following Cauchy problem:
\begin{eqnarray}
	\Lhat_H \Phat_1(\tau,k,z)
		&=& \Ahat \Ghat(\tau,k,z) \hhat(k), \label{eq:P1hatPDE} \\
	\Lhat_H &=& 	-\frac{\d}{\d \tau} + \frac{1}{2} z \left(-k^2 + i k \right) + \frac{1}{2} \sigma^2 z \frac{\d^2}{\d z^2} \nn \\
					& & 	+ \left( \kappa \theta - \left(\kappa + \rho \sigma i k \right) z  \right) \frac{\d}{\d z} \,,\nn \\ 
	\Ahat	
		&=& z  \left( V_1 \frac{\d}{\d z} \left(-k^2+ ik \right) + V_2 \frac{\d^2}{\d z^2} \left( -ik \right) \right. \nn \\
    & &  \left. + V_3 \left( ik^3 + k^2 \right) + V_4 \frac{\d}{\d z}\left( -k^2 \right) \right) ,\nn \\
  \Phat_1(0,k,z)
  	&=& 0 \label{eq:P1hatBC}.
\end{eqnarray}
Now, an ansatz: we suppose that $\Phat_1(\tau,k,z)$ can be written as
\begin{eqnarray}
	\Phat_1(\tau,k,z)=\left(\kappa \theta \fhat_0(\tau,k)+z \fhat_1(\tau,k))\right)\Ghat(\tau,k,z)\hhat(k) . \label{eq:P1Ansatz}
\end{eqnarray} 
We substitute (\ref{eq:P1Ansatz}) into (\ref{eq:P1hatPDE}) and (\ref{eq:P1hatBC}).  After a good deal of algebra (and in particular, making use of (\ref{eq:ODEforC}) and (\ref{eq:ODEforD})), we find that $\fhat_0(\tau,k)$ and $\fhat_1(\tau,k)$ satisfy the following system of ODE's:
\begin{eqnarray}
	\frac{d \fhat_1}{d \tau}(\tau,k) 	&=& 	a(\tau,k) \widehat{f}_1(\tau,k) + b(\tau,k), 	\label{eq:f1hatODE} \\
	\fhat_1(0,k)											&=&		0 , \label{eq:f1hatBC} \\
	\frac{d \fhat_0}{d \tau}(\tau,k) 	&=& 	\fhat_1(\tau,k),	\label{eq:f0hatODE} \\
	\fhat_0(0,k) 											&=& 	0,			\label{eq:f0hatBC} \\
	a(\tau,k)												  &=& 	\sigma^2 D(\tau,k) -\left(\kappa + \rho \sigma i k \right), \nn \\
	b(\tau,k)												  &=&  -  \left( V_1 D(\tau,k) \left(-k^2+ ik \right) + V_2 D^2(\tau,k) \left( -ik \right) \right. \nn \\
		        											  &	&		\left. + V_3 \left(ik^3 + k^2\right) + V_4 D(\tau,k)\left( -k^2 \right) \right) ,\nn 
\end{eqnarray}
where $D(\tau,k)$ is given by equation (\ref{eq:D}).
\par
Equations (\ref{eq:f1hatODE}--\ref{eq:f0hatBC}) can be solved analytically (to the extent that their solutions can be written down in integral form).  The solutions for $\fhat_0(\tau,k)$ and $\fhat_1(\tau,k)$, along with the final solution for $P_1(t,x,z)$, are given by (\ref{eq:P1solution}--\ref{eq:b}).

\section{Moment Estimate for $Y_t$} \label{appendixbound}
In this section we will derive a moment estimate for $Y_t$, whose dynamics under the pricing measure are given by equations (\ref{eq:OU}, \ref{eq:CIR}, \ref{eq:rhoYZ}).  Specifically, we will show that for all $\alpha\in (1/2,1)$ there exists a constant, $C$ (which depends on $\alpha$ but independent of $\eps$), such that $\E |Y_t|  \leq C \, \eps^{\alpha-1}$.

We will begin by estabilshing some notation.  First we define a continuous, strictly increasing, non-negative process, $\beta_t$, as
\begin{align*}
	\beta_t &:= \int_0^t Z_s ds .
\end{align*}
Next, we note that $W_t^y$ may be decompoased as
\begin{align}
	W_t^y 	&= \rho_{yz} W_t^z + \sqrt{1-\rho_{yz}^2} W_t^\perp ,	\label{eq:Wdecomposition} 
\end{align}
where $W_t^\perp$ is a Brownian motion which is independent of $W_t^z$.  Using equations (\ref{eq:Yexplicit}) and (\ref{eq:Wdecomposition}) we derive
\begin{align*}
	|Y_t|
		&\leq		C_1 + \frac{C_2}{\sqrt{\eps}} \left[ e^{\frac{-1}{\eps}\beta_t} \left| 	\int_0^t e^{\frac{1}{\eps}\beta_s} \sqrt{Z_s} dW_s^z	\right|
																								+e^{\frac{-1}{\eps}\beta_t} \left| \int_0^t e^{\frac{1}{\eps}\beta_s} \sqrt{Z_s} dW_s^\perp \right|	\right] ,
\end{align*}
where $C_1$ and $C_2$ are constants.  We will focus on bounding the first moment of the second stochastic integral.  We have:
\begin{align*}
	\frac{1}{\eps} \E \left[	\left( e^{\frac{-1}{\eps}\beta_t} \int_0^t e^{\frac{1}{\eps}\beta_s} \sqrt{Z_s} dW_s^\perp \right)^2	\right]
	&= \frac{1}{\eps}
	\E \left[	e^{-2 \beta_t/\eps} \left. \E \left[ \left( \int_0^t e^{\beta_s/\eps} \sqrt{Z_s} dW_s^\perp \right)^2 \right| \beta_t	\right]	\right]\\
	&= \frac{1}{\eps} \E \left[	e^{-2 \beta_t/\eps} \left. \E \left[ \int_0^t e^{2 \beta_s/\eps} Z_s ds \right| \beta_t	\right]	\right]\\
	&= \frac{1}{\eps} \E\left[	e^{-2 \beta_t/\eps} \left. \E \left[ \int_0^t e^{2 \beta_s/\eps} d\beta_s \right| \beta_t	\right]	\right]\\
	&= \frac{1}{\eps} \E\left[	e^{-2 \beta_t/\eps} \frac{\eps}{2}\left( e^{2 \beta_t/\eps} -1 \right)	\right]\\
	&=	\frac{1}{2} \E \left[ 1 - e^{-\frac{2}{\eps}\beta_t} \right]	\leq \frac{1}{2} .
\end{align*}
Then, by the Cauchy-Schwarz inequality, we see that
\begin{align*}
	\frac{1}{\sqrt{\eps}} \E \left[ e^{\frac{-1}{\eps}\beta_t}\left| \int_0^t e^{\frac{1}{\eps}\beta_s} \sqrt{Z_s} dW_s^\perp \right| \right]
		&\leq \frac{1}{\sqrt{2}}	.
\end{align*}
What remains is to bound the first moment of the other stochastic integral,
\begin{align*}
	A &:=	\frac{1}{\sqrt{\eps}} \E \left[ e^{\frac{-1}{\eps}\beta_t}\left| \int_0^t e^{\frac{1}{\eps}\beta_s} \sqrt{Z_s} dW_s^z \right| \right]	.
\end{align*}
Naively, one might try to use the Cauchy-Schwarz inequality in the following manner
\begin{align*}
	A &\leq	\frac{1}{\sqrt{\eps}} \sqrt{\E \left[ e^{-2 \beta_t/\eps}\right]} \sqrt{\E \left[ \int_0^t e^{2 \beta_s/\eps} Z_s ds \right]}	\\
		&=\frac{1}{\sqrt{\eps}} \sqrt{\E \left[ e^{-2 \beta_t/\eps}\right]} \sqrt{\E \left[ \frac{\eps}{2}\left( e^{2 \beta_t/\eps}\right) \right]}.
\end{align*}
However, this approach does not work, since $\E\left[ e^{2 \beta_t/\eps}\right] \rightarrow \infty$ as $\eps \rightarrow 0$.  Seeking a more refined approach of bounding $A$, we note that
\begin{align*}
	\frac{1}{\sqrt{\eps}} e^{\frac{-1}{\eps}\beta_t}\int_0^t e^{\frac{1}{\eps}\beta_s} \sqrt{Z_s} dW_s^z
	&=	\frac{1}{\sigma \sqrt{\eps}}e^{-\frac{1}{\eps}\beta_t}(Z_t-z)
			-	\frac{\kappa}{\sigma \sqrt{\eps}} e^{-\frac{1}{\eps}\beta_t}\int_0^t e^{\frac{1}{\eps}\beta_s}(\theta - Z_s) ds	\\
	&\quad		+	\frac{1}{\sigma \eps^{3/2}} e^{-\frac{1}{\eps}\beta_t}\int_0^t e^{\frac{1}{\eps}\beta_s} Z_s(Z_t-Z_s) ds	,
\end{align*}
which can be derived by replacing $t$ by $s$ in equation (\ref{eq:CIR}), multiplying by $e^{\beta_s/\eps}$, integrating the result from $0$ to $t$ and using $Z_s^2 = Z_t Z_s - Z_s (Z_t - Z_s)$ and $\int_0^t e^{-(\beta_t-\beta_s)/\eps} Z_s ds = \eps (1-e^{-\beta_t/\eps})$.  From the equation above, we see that
\begin{align}
	A	&\leq		\frac{1}{\sigma \sqrt{\eps}}\E \left[ e^{-\frac{1}{\eps}\beta_t} |Z_t-z| 	\right]
					+	\frac{\kappa}{\sigma \sqrt{\eps}} \E \left[ e^{-\frac{1}{\eps}\beta_t} \left| \int_0^t e^{\frac{1}{\eps}\beta_s}(\theta - Z_s) ds \right|	\right] \nn\\
	&\quad	+	\frac{1}{\sigma \eps^{3/2}} \E \left[ e^{-\frac{1}{\eps}\beta_t} \left| \int_0^t e^{\frac{1}{\eps}\beta_s} Z_s(Z_t - Z_s) ds	\right| \right]
	\label{eq:Abound} .
\end{align}
At this point, we need the moment generating function of $(Z_t,\beta_t)$.  From \cite{ll}, we have
\begin{align}
	\E \left[ e^{-\lambda Z_t - \mu \beta_t} \right]	&=	e^{-\kappa \theta \phi_{\lambda,\mu}(t)-z \psi_{\lambda,\mu}(t)},\label{eq:MomentGenFunc}	\\
	\phi_{\lambda,\mu}(t) &=
		\frac{-2}{\sigma^2}\log\left[\frac{2 \gamma  e^{(\gamma +\kappa )t/2} }
		{\lambda  \sigma ^2\left(e^{ \gamma  t}-1\right)+\gamma -\kappa +e^{\gamma  t} (\gamma +\kappa ) }\right] , \nn \\
	\psi_{\lambda,\mu}(t) &=
		\frac{\lambda \left(\gamma +\kappa +e^{\text{$\gamma $} t} (\gamma -\kappa )\right) +2 \mu \left(e^{\gamma  t}-1\right) }
		{\lambda  \sigma ^2\left(e^{ \gamma  t}-1\right)+\gamma -\kappa +e^{\gamma  t} (\gamma +\kappa )} , \nn	\\
	\gamma	&=	\sqrt{\kappa^2 + 2 \sigma^2 \mu} \nn .
\end{align}
Now, let us focus on the first term in equation (\ref{eq:Abound}).  Using Cauchy-Schwarz, we have
\begin{align*}
	\frac{1}{\sigma \sqrt{\eps}}\E \left[ e^{-\beta_t/\eps} |Z_t-z| 	\right]
		&\leq		\frac{1}{\sigma \sqrt{\eps}} \sqrt{\E \left[e^{-2 \beta_t/\eps}	\right] }  \sqrt{\E \left[ |Z_t-z|^2 \right] }	.
\end{align*}
From equation (\ref{eq:MomentGenFunc}) one can verify
\begin{align*}
	\E \left[ |Z_t-z|^2 \right] &\leq C_3	, \\
	\E \left[e^{-2 \beta_t/\eps}	\right] &= e^{-\kappa \theta \phi_{0,2/\eps}(t)-z \phi_{0,2/\eps}(t)} \sim e^{C_4/\sqrt{\eps}} ,
\end{align*}
where $C_3$ and $C_4$ are constants.  Since $\frac{1}{\sqrt{\eps}}e^{C_4/\sqrt{\eps}} \rightarrow 0$ as $\eps \rightarrow 0$ we see that
\begin{align*}
	\frac{1}{\sigma \sqrt{\eps}}\E \left[e^{-\beta_t/\eps} |Z_t-z| 	\right] &\leq C_5	,
\end{align*}
for some constant $C_5$.

We now turn out attention to the second term in equation (\ref{eq:Abound}).  We have
\begin{align*}
	&\frac{\kappa}{\sigma \sqrt{\eps}} \E \left[ \left| \int_0^t e^{-\frac{1}{\eps}(\beta_t-\beta_s)}(\theta - Z_s) ds \right|	\right]	\\
	&\leq	\frac{\kappa}{\sigma \sqrt{\eps}} \E \left[  \int_0^t e^{-\frac{1}{\eps}(\beta_t-\beta_s)}Z_s ds 	\right]
					+	\frac{\kappa \theta}{\sigma \sqrt{\eps}} \E \left[  \int_0^t e^{-\frac{1}{\eps}(\beta_t-\beta_s)}ds	\right]	\\
	&\leq	\frac{\kappa}{\sigma \sqrt{\eps}} \E \left[  \int_0^t e^{-\frac{1}{\eps}(\beta_t-\beta_s)}d\beta_s 	\right]
					+	\frac{\kappa \theta}{\sigma \sqrt{\eps}} \E \left[  \int_0^t e^{-\frac{1}{\eps}(\beta_t-\beta_s)}ds 	\right]	\\	
	&\leq	\frac{\kappa}{\sigma \sqrt{\eps}} \E \left[  \eps \left( 1 - e^{-\beta_t/\eps} \right) 	\right]
					+	\frac{\kappa \theta}{\sigma \sqrt{\eps}} \E \left[  \int_0^t e^{-\frac{1}{\eps}(\beta_t-\beta_s)}ds 	\right]	\\
	&\leq C_6
					+	\frac{\kappa \theta}{\sigma \sqrt{\eps}} \int_0^t \E \left[ e^{-\frac{1}{\eps}(\beta_t-\beta_s)} 	\right] ds ,
\end{align*}
for some constant $C_6$.  To bound the remaining integral we calculate
\begin{align}
	\E \left[ e^{-\frac{1}{\eps}(\beta_t-\beta_s)} 	\right]
		&=	\E \left[ \left. \E \left[ e^{-\frac{1}{\eps}(\beta_t-\beta_s)} \right| Z_s \right] 	\right]	\nn \\
		&=	\E \left[ e^{-\kappa \theta \phi_{0,1/\eps}(t-s)-Z_s \psi_{0,1/\eps}(t-s)} \right]	\nn \\
		&=	\exp \Big( -\kappa \theta \phi_{0,1/\eps}(t-s)-\kappa \theta \phi_{\bar{\psi},0}(s) - z \psi_{\bar{\psi},0}(s) \Big)	\label{eq:BetaMoment} \\
	\bar{\psi}(s)
		&:=	\psi_{0,1/\eps}(t-s) . \nn
\end{align}
Using the fact that $\phi_{\lambda,\mu}(t),\psi_{\lambda,\mu}(t)>0$ for any $\lambda,\mu,t>0$, we see that
\begin{align*}
	\E \left[ e^{-\frac{1}{\eps}(\beta_t-\beta_s)} 	\right]
		&\leq	\exp \Big( - z \psi_{\bar{\psi},0}(s) \Big)	.
\end{align*}
Hence
\begin{align}
	\int_0^t \E \left[ e^{-\frac{1}{\eps}(\beta_t-\beta_s)} 	\right] ds
		&\leq 		\int_0^{t-\eps^\alpha} \exp \Big( - z \psi_{\bar{\psi},0}(s) \Big) ds
					+	\int_{t-\eps^\alpha}^t \exp \Big( - z \psi_{\bar{\psi},0}(s) \Big) ds	\nn \\
		&=: I_1 + I_2	,	\label{eq:I2I1}
\end{align}
where $\alpha \in (1/2,1)$.  Using again $\psi_{\lambda,\mu}(t)>0$ we deduce $\psi_{\bar{\psi},0}(s)>0$ and therefore
\begin{align}
I_1
	&\leq \eps^\alpha.
	\end{align}
As for $I_1$, we claim
\begin{align}
I_1
	&\leq C_7 \exp \left( - C_8 \eps^{\alpha-1} \right)	, \label{eq:I2bound}
\end{align}
which is equivalent to showing there exists a constant $C$ such that
\begin{align}
\psi_{\bar{\psi},0}(s)
	&\geq C \eps^{\alpha-1}	\label{eq:claim2}
\end{align}
for all $s \in [0,t-\eps^\alpha]$.  To prove this claim, we note that for small $\eps$
\begin{align*}
\bar{\psi}(s)
	&=		\psi_{0,1/\eps}(t-s)
	\sim	\frac{\sigma \sqrt{2}}{\sqrt{\eps}}\left(\frac
	{\exp\left[ \frac{\sigma \sqrt{2}}{\sqrt{\eps}}(t-s)\right]-1}
	{\exp\left[ \frac{\sigma \sqrt{2}}{\sqrt{\eps}}(t-s)\right]+1}\right)	,
\end{align*}
where we have used $\gamma = \sqrt{\kappa^2+2\sigma^2/\eps}\sim \sigma\sqrt{2}/\sqrt{\eps}$.  A direct computation shows that $\bar{\psi}(s)$ is a strictly decreasing in $s$ with
\begin{align*}
\bar{\psi}(t-\eps^\alpha)=\psi_{0,1/\eps}(\eps^\alpha)
	&\sim \sigma^2	\eps^{\alpha-1}	.
\end{align*}
Now, we note that $\psi_{\bar{\psi},0}(s)$ is given by
\begin{align*}
\psi_{\bar{\psi},0}(s)
	&=	\frac{2 \kappa \bar{\psi}(s)}{\sigma^2\left(e^{\kappa s}-1\right)\bar{\psi}(s) + 2 \kappa e^{\kappa s}}
	=		\frac{2 \kappa }{\sigma^2\left(e^{\kappa s}-1\right) + 2 \kappa e^{\kappa s}/\bar{\psi}(s)}.
\end{align*}
Since $e^{\kappa s}<e^{\kappa t}$, and since, at worst, $\bar{\psi}(s)\sim \sigma^2\eps^{\alpha-1}$, we conclude that there exists a constant $C$ such that (\ref{eq:claim2}), and therefore (\ref{eq:I2bound}), hold.  Hence, using equation (\ref{eq:I2I1}-\ref{eq:I2bound}), we have
\begin{align*}
	\int_0^t \E \left[ e^{-\frac{1}{\eps}(\beta_t-\beta_s)} 	\right] ds
		&= 		\int_0^{t-\eps^\alpha} \E \left[ e^{-\frac{1}{\eps}(\beta_t-\beta_s)} 	\right] ds
					+	\int_{t-\eps^\alpha}^t \E \left[ e^{-\frac{1}{\eps}(\beta_t-\beta_s)} 	\right] ds	\\
		&\leq	C_7 e^{-C_8 \eps^{\alpha-1}} +  \eps^\alpha	.
\end{align*}
This implies that there exists a constant $C_{9}$ such that for any $\alpha \in (1/2,1)$
\begin{align*}
	\frac{\kappa \theta}{\sigma \sqrt{\eps}} \int_0^t \E \left[ e^{-\frac{1}{\eps}(\beta_t-\beta_s)} 	\right] ds
		&\leq C_{9}	.
\end{align*}
Having established a uniform bound on the first two terms in equation (\ref{eq:Abound}), we turn our attention toward the third and final term.  For $\alpha\in(1/2,1)$ we have
\begin{align*}
	\frac{1}{\sigma \eps^{3/2}} \E \left[ \left| \int_0^t e^{-\frac{1}{\eps}(\beta_t-\beta_s)} Z_s(Z_t - Z_s) ds	\right| \right]
	&\leq	\frac{1}{\sigma \eps^{3/2}} \E \left[  \int_0^t e^{-\frac{1}{\eps}(\beta_t-\beta_s)} Z_s|Z_t - Z_s| ds \right]	\\
	&=	\frac{1}{\sigma \eps^{3/2}} \E \left[  \int_0^{t-\eps^\alpha} e^{-\frac{1}{\eps}(\beta_t-\beta_s)} Z_s|Z_t - Z_s| ds \right]	\\
	&\quad		+	\frac{1}{\sigma \eps^{3/2}} \E \left[  \int_{t-\eps^\alpha}^t e^{-\frac{1}{\eps}(\beta_t-\beta_s)} Z_s|Z_t - Z_s| ds \right] .
\end{align*}
For the integral from $0$ to $(t-\eps^\alpha)$ we compute
\begin{align*}
		&\frac{1}{\sigma \eps^{3/2}} \E \left[  \int_0^{t-\eps^\alpha} e^{-\frac{1}{\eps}(\beta_t-\beta_s)} Z_s|Z_t - Z_s| ds \right]	\\
		&\leq \frac{1}{\sigma \eps^{3/2}} \sqrt{\E \left[ \left( \sup_{0 \leq s \leq t} Z_s|Z_t-Z_s| \right)^2 \right]}
					\sqrt{\E \left[  \int_0^{t-\eps^\alpha} e^{-\frac{1}{\eps}(\beta_t-\beta_s)}ds \right]}	\\
		&\leq \frac{1}{\eps^{3/2}} C_{10} e^{-C_{11} \eps^{\alpha-1}}	\leq C_{12} ,
\end{align*}
for some constants $C_{10}$, $C_{11}$ and $C_{12}$.  For the integral from $(t-\eps^\alpha)$ to $t$ we have
\begin{align*}
	&\frac{1}{\sigma \eps^{3/2}} \E \left[  \int_{t-\eps^\alpha}^t e^{-\frac{1}{\eps}(\beta_t-\beta_s)} Z_s|Z_t - Z_s| ds \right]	\\
	&\leq \frac{1}{\sigma \eps^{3/2}}
				\E \left[ \sup_{t-\eps^\alpha \leq s \leq t}|Z_t-Z_s|  \int_{t-\eps^\alpha}^t e^{-\frac{1}{\eps}(\beta_t-\beta_s)} Z_s ds \right]	\\
	&=		\frac{1}{\sigma \eps^{3/2}}
				\E \left[ \sup_{t-\eps^\alpha \leq s \leq t}|Z_t-Z_s| \, \eps \, (1-e^{-\beta_{(t-\eps^\alpha)/\eps}}) \right]	\\
	&\leq	\frac{1}{\sigma \eps^{1/2}}	\E \left[ \sup_{t-\eps^\alpha \leq s \leq t}|Z_t-Z_s| \right]		\leq 	C_{13} \eps^{\alpha-1} ,
\end{align*}
for some constant $C_{13}$.  With this result, we have established that for all $\alpha\in (1/2,1)$ there exists a constant, $C$, such that $\E |Y_t|  \leq C \, \eps^{\alpha-1}$.

\section{Numerical Computation of Option Prices} \label{sec:numerical}
The formulas (\ref{eq:PHsolution}) and (\ref{eq:P1solution}) for $P_H(t,x,z)$  and $P_1(t,x,z)$  cannot be evaluated analytically.  Therefore, in order for these formulas to be useful, an efficient and reliable numerical integration scheme is needed.  Unfortunately, numerical evaluation of the integral in (\ref{eq:PHsolution}) is notoriously difficult.  And, the double and triple integrals that appear in (\ref{eq:P1solution}) are no easier to compute.  In this section, we point out some of the difficulties associated with numerically evaluating these expressions, and show how these difficulties can be addressed.  We begin by establishing some notation.
\begin{eqnarray*}
	P^{\eps}(t,x,z) & \sim &  P_H(t,x,z)+\sqrt{\eps} P_1(t,x,z), \\
		&=&  	\frac{e^{-r\tau}}{2 \pi}
					\int_\R e^{-ikq} \left(1 + \sqrt{\eps} \left( \kappa \theta \fhat_0(\tau,k) + z \fhat_1(\tau,k) \right) \right)  \\
		& &		\times \Ghat(\tau,k,z) \hhat(k) dk,  \\
		&=&  	\frac{e^{-r\tau}}{2 \pi}
				\left( P_{0,0}(t,x,z) + \kappa \theta \sqrt{\eps}P_{1,0}(t,x,z) + z \sqrt{\eps} P_{1,1}(t,x,z) \right),
\end{eqnarray*}
where we have defined
\begin{eqnarray}
	P_{0,0}(t,x,z) 	&:=& \int_\R e^{-ikq} \Ghat(\tau,k,z) \hhat(k) dk \label{eq:P00}, \\
	P_{1,0}(t,x,z) 	&:=& \int_\R e^{-ikq} \fhat_0(\tau,k) \Ghat(\tau,k,z) \hhat(k) dk \label{eq:P10}, \\
	P_{1,1}(t,x,z)	&:=& \int_\R e^{-ikq} \fhat_1(\tau,k) \Ghat(\tau,k,z) \hhat(k) dk \label{eq:P11} .
\end{eqnarray}
As they are written, (\ref{eq:P00}), (\ref{eq:P10}) and (\ref{eq:P11}) are general enough to  accomodate any European option.  However, in order to make  progress, we now specify an option payoff.  We will limit ourself to the case of an European call, which has payoff $h(x)=(x-K)^+$.  Extension to other European options is straightforward.
\par
We remind the reader that $\hhat(k)$ is the Fourier transform of the option payoff, expressed as a function of $q=r(T-t)+\log(x)$.  For the case of the European call, we have:
\begin{eqnarray}
	\hhat(k)	&=& \int_\R e^{i k q} (e^q-K)^+ dq  \label{eq:hhatintegral}
						=	\frac{K^{1+ik}}{ik-k^2}.	
\end{eqnarray}
We note that (\ref{eq:hhatintegral}) will not converge unless the imaginary part of $k$ is greater than one.  Thus, we decompose $k$ into its real and imaginary parts, and impose the following condition on the imaginary part of $k$.
\begin{eqnarray}
	k		&=&	k_r + i k_i ,\nn\\
	k_i &>&	1.\label{eq:imaginary}
\end{eqnarray}
When we integrate over $k$ in  (\ref{eq:P00}), (\ref{eq:P10}) and (\ref{eq:P11}), we hold $k_i>1$ fixed, and integrate $k_r$ over $\R$.

\subsection*{Numerical Evaluation of $P_{0,0}(t,x,z)$}
We rewrite (\ref{eq:P00}) here, explicitly using expressions (\ref{eq:Ghat}) and (\ref{eq:hhatintegral}) for $\Ghat(\tau,k,z)$ and $\hhat(k)$ respectively.
\begin{eqnarray}
	P_{0,0}(t,x,z) &=& \int_\R e^{-ikq} e^{C(\tau,k)+z D(\tau,k)} \frac{K^{1+ik}}{ik-k^2} dk_r. \label{eq:P00explicit}
\end{eqnarray}
In order for any numerical integration scheme to work, we must verify the continuity of the integrand in (\ref{eq:P00explicit}).  
First, by (\ref{eq:imaginary}), the poles at $k=0$ and $k=i$ are avoided.
The only other worrisome term in the integrand of (\ref{eq:P00explicit}) is $e^{C(\tau,k)}$, which may be discontinuous due to the presence of the $\log$ in $C(\tau,k)$.
\par
We recall that any $\zeta \in \mathbb{C}$ can be represented in polar notation as $\zeta = r \exp(i \theta)$, where $\theta \in [-\pi,\pi)$.  In this notation, $\log \zeta = \log r + i \theta$.  Now, suppose we have a map $\zeta(k_r):\mathbb{R} \rightarrow \mathbb{C}$.  We see that whenever $\zeta(k_r)$ crosses the negative real axis, $\log \zeta(k_r)$ will be discontinuous (due to $\theta$ jumping from $-\pi$ to $\pi$ or from $\pi$ to $-\pi$).  Thus, in order for $\log \zeta(k_r)$ to be continuous, we must ensure that $\zeta(k_r)$ does not cross the negative real axis.
\par
We now return our attention to $C(\tau,k)$.  We note that $C(\tau,k)$ has two algebraically equivalent representations,  (\ref{eq:C}) and the following representation:
\begin{eqnarray}
	C(\tau,k)				&=& 	\frac{\kappa \theta}{\sigma^2} \left( \left( \kappa + \rho i k \sigma - d(k) \right) \tau 
												- 2 \log \zeta(\tau,k) \right), \label{eq:C2} \\
	\zeta (\tau,k) 	&:=&	\frac{e^{-\tau d(k)}/g(k)-1}{1/g(k)-1}.	 \label{eq:zeta}
\end{eqnarray}
It turns out that, under most reasonable conditions, $\zeta(\tau,k)$ does not cross the negative real axis \cite{LordKahl2006}.  As such, as one integrates over $k_r$, no discontinuities will arise from the $\log \zeta(\tau,k)$ which appears in (\ref{eq:C2}).  Therefore, if we use expression (\ref{eq:C2}) when evaluating (\ref{eq:P00explicit}), the integrand will be continuous.

\subsection*{Numerical Evaluation of $P_{1,1}(t,x,z)$ and $P_{1,0}(t,x,z)$}
 The integrands in (\ref{eq:P11}) and (\ref{eq:P10}) are identical to that of (\ref{eq:P00}), except for the additional factor of $\fhat_1(\tau,k)$. Using equation (\ref{eq:f1hat}) for $\fhat_1(\tau,k)$ we have the following expression for $P_{1,1}(t,x,z)$:
\begin{eqnarray}
	\lefteqn{	P_{1,1}(t,x,z) } \nn \\
	&=&	\int_\R e^{-ikq} \left( \int_0^\tau b(s,k) e^{A(\tau,k,s)} ds \right) e^{C(\tau,k)+z D(\tau,k)} \frac{K^{1+ik}}{ik-k^2} dk_r \nn  \\
	&=&	\int_0^\tau \int_\R e^{-ikq} b(s,k) e^{A(\tau,k,s)+C(\tau,k)+zD(\tau,k)} \frac{K^{1+ik}}{ik-k^2} dk_r ds  . \label{eq:P11explicit}
\end{eqnarray} 
Similarly:
\begin{eqnarray}
	\lefteqn{	P_{1,0}(t,x,z) } \nn \\
	&=&	\int_0^\tau \int_0^t \int_\R e^{-ikq} b(s,k) e^{A(t,k,s)+C(\tau,k)+zD(\tau,k)} \frac{K^{1+ik}}{ik-k^2} dk_r ds dt .
			\label{eq:P10explicit}
\end{eqnarray}
We already know, from our analysis of $P_{0,0}(t,x,z)$, how to deal with the $\log$ in $C(\tau,k)$.  It turns out that the $\log$ in $A(\tau,k,s)$ can be dealt with in a similar manner.  Consider the following representation for $A(\tau,k,s)$, which is algebraically equivalent to expression (\ref{eq:A}):
\begin{eqnarray}
	A(\tau,k,s)	&=& 	\left( \kappa + \rho i k \sigma + d(k) \right) \left(\frac{1-g(k)}{d(k) g(k)} \right)  \nn \\
				& &		\times\left( d(k)(\tau - s) + \log \zeta (\tau,k) - \log \zeta (s,k) \right) \nn \\
				& &		+ d(k) (\tau - s) , \label{eq:A2}
\end{eqnarray}
where $\zeta(\tau,k)$ is defined in (\ref{eq:zeta}).  As expressed in (\ref{eq:A2}), $A(\tau,k,s)$ is, under most reasonable conditions, a continuous function of $k_r$.  Thus, if we use (\ref{eq:A2}) when numerically evaluating (\ref{eq:P11explicit}) and (\ref{eq:P10explicit}), their integrands will be continuous.

\subsection*{Transforming the Domain of Integration}    
Aside from using equations (\ref{eq:C2}) and (\ref{eq:A2}) for $C(\tau,k)$ and $A(\tau,k,s)$, there are a few other tricks we can use to facilitate the numerical evaluation of (\ref{eq:P00explicit}), (\ref{eq:P10explicit}), and (\ref{eq:P11explicit}).  Denote by $I_0(k)$ and $I_1(k,s)$ the integrands appearing in (\ref{eq:P00explicit}), (\ref{eq:P11explicit}) and (\ref{eq:P10explicit}).
\begin{eqnarray*}
	P_{0,0} &=& \int_\R I_0(k) dk_r, \\
	P_{1,1} &=& \int_0^\tau \int_\R I_1(k,s) dk_r ds, \\
	P_{1,0} &=& \int_0^\tau \int_0^t \int_\R I_1(k,s) dk_r ds dt.
\end{eqnarray*}
First, we note that the real and imaginary parts of  $I_0(k)$ and $I_1(k,s)$ are even and odd functions of $k_r$ respectively.  As such, instead of integrating in $k_r$ over $\R$, we can integrate in $k_r$ over $\R_+$, drop the imaginary part, and multiply the result by $2$.
\par
Second, numerically integrating in $k_r$ over $\R_+$ requires that one arbitrarily truncate the integral at some $k_{cutoff}$.  Rather than doing this, we can make the following variable transformation, suggested by \cite{JackelKahl2005}:
\begin{eqnarray}
	k_r					&=&		\frac{-\log u}{C_{\infty}} ,\nn \\
	C_{\infty}	&:=&	\frac{\sqrt{1-\rho^2}}{\sigma}(z+\kappa \theta \tau) . \label{eq:Cinfty}
\end{eqnarray}
Then, for some arbitrary $I(k)$ we have
\begin{eqnarray*}
	\int_0^\infty I(k) dk_r &=& \int_0^1 I \left( \frac{-\log u}{C_{\infty}} + i k_i \right) \frac{1}{u C_{\infty}} du .
\end{eqnarray*}
Thus, we avoid having to establish a cutoff value, $k_{cutoff}$ (and avoid the error that comes along with doing so).
\par
Finally, evaluating (\ref{eq:P10explicit}) requires that one integrates over the triangular region parameterized by $0 \leq s \leq t \leq \tau$.  Unfortunately, most numerical integration packages only facilitate integration over a rectangular region.  We can overcome this difficulty by performing the following transformation of variables:
\begin{eqnarray*}
	s		&=& t v, \\
	ds	&=&	t dv .
\end{eqnarray*}  
Then, for some arbitrary $I(s)$ we have
\begin{eqnarray}
	\int_0^\tau \int_0^t I(s) ds dt	&=&	\int_0^\tau \int_0^1 I(t v) t dv dt.
\end{eqnarray}
Pulling everything together we obtain:
\begin{eqnarray*}
	P_{0,0} &=& 2 Re \int_0^1 I_0\left( \frac{-\log u}{C_{\infty}} + i k_i \right) \frac{1}{u C_{\infty}} du, \\
	P_{1,1} &=& 2 Re \int_0^\tau \int_0^1 I_1\left( \frac{-\log u}{C_{\infty}} + i k_i,s \right) \frac{1}{u C_{\infty}} du ds, \\
	P_{1,0} &=& 2 Re \int_0^\tau \int_0^1 \int_0^1 I_1\left( \frac{-\log u}{C_{\infty}} + i k_i, t v \right) \frac{t}{u C_{\infty}} du dv dt ,
\end{eqnarray*}
where $C_{\infty}$ is given by (\ref{eq:Cinfty}).  These three changes allow one to efficiently and accurately numerically evaluate (\ref{eq:P00explicit}), (\ref{eq:P11explicit}) and (\ref{eq:P10explicit}).

Numerical tests show that for strikes ranging from $0.5$ to $1.5$ the spot price, and for expirations ranging from $3$ months to $3$ years, it takes roughly 100 times longer to calculate a volatility surface using the multi-scale model than it does to calculate the same surface using the Heston model.

\subsection*{Acknowledgment} 

The authors would like to thank Ronnie Sircar and Knut S{\o}lna for earlier discussions on the model studied in this paper. They also thank two anonymous referees for their suggestions that greatly helped improve the paper.

\bibliographystyle{plain}
\bibliography{FouqueHestonBibtex1}

\begin{thebibliography}{10}

\bibitem{alizadeh2001}
Sassan Alizadeh, Michael~W. Brandt, and Francis~X. Diebold.
\newblock {Range-Based Estimation of Stochastic Volatility Models}.
\newblock {\em SSRN eLibrary}, 2001.

\bibitem{anderson}
Torben~G. Andersen and Tim Bollerslev.
\newblock Intraday periodicity and volatility persistence in financial markets.
\newblock {\em Journal of Empirical Finance}, 4(2-3):115--158, June 1997.

\bibitem{chernov2003}
Mikhail Chernov, A.~Ronald~Gallant, Eric Ghysels, and George Tauchen.
\newblock Alternative models for stock price dynamics.
\newblock {\em Journal of Econometrics}, 116(1-2):225--257, 2003.

\bibitem{cfps}
Peter Cotton, Jean-Pierre Fouque, George Papanicolaou, and Ronnie Sircar.
\newblock Stochastic volatility corrections for interest rate derivatives.
\newblock {\em Mathematical Finance}, 14(2), 2004.

\bibitem{engle}
Robert~F. Engle and Andrew~J. Patton.
\newblock What good is a volatility model?
\newblock 2008.

\bibitem{fiorentini2002}
Gabriele Fiorentini, Angel Leon, and Gonzalo Rubio.
\newblock Estimation and empirical performance of {H}eston's stochastic
  volatility model: the case of a thinly traded market.
\newblock {\em Journal of Empirical Finance}, 9(2):225--255, March 2002.

\bibitem{fouque}
Jean-Pierre Fouque, George Papanicolaou, and Ronnie Sircar.
\newblock {\em Derivatives in Financial Markets with Stochastic Volatility}.
\newblock Cambridge University Press, 2000.

\bibitem{fouque2}
Jean-Pierre Fouque, George Papanicolaou, Ronnie Sircar, and Knut Solna.
\newblock Short time-scale in {S}\&{P} 500 volatility.
\newblock {\em The Journal of Computational Finance}, 6(4), 2003.

\bibitem{proof}
Jean-Pierre Fouque, George Papanicolaou, Ronnie Sircar, and Knut Solna.
\newblock Singular perturbations in option pricing.
\newblock {\em SIAM J. Applied Mathematics}, 63(5):1648--1665, 2003.

\bibitem{gatheral2}
Jim Gatheral.
\newblock Modeling the implied volatility surface.
\newblock In {\em Global Derivatives and Risk Management, Barcelona}, May 2003.

\bibitem{gatheral}
Jim Gatheral.
\newblock {\em The Volatility Surface: a Practitioner's Guide}.
\newblock John Wiley and Sons, Inc., 2006.

\bibitem{heston}
Steven Heston.
\newblock {A closed-form solution for options with stochastic volatility with
  applications to bond and currency options}.
\newblock {\em Rev. Financ. Stud.}, 6(2):327--343, 1993.

\bibitem{hillebrand}
Eric Hillebrand.
\newblock Overlaying time scales and persistence estimation in {GARCH}(1,1)
  models.
\newblock Econometrics 0301003, EconWPA, January 2003.

\bibitem{JackelKahl2005}
Peter Jackel and Christian Kahl.
\newblock {Not-so-complex logarithms in the {H}eston Model}.
\newblock {\em Wilmott}, 2005.

\bibitem{ll}
Damien Lamberton and Bernard Lapeyre.
\newblock {\em Introduction to Stochastic Calculus Applied to Finance}.
\newblock Chapman \& Hall, 1996.

\bibitem{lebaron}
Blake~D. Lebaron.
\newblock {Stochastic Volatility as a Simple Generator of Financial Power-Laws
  and Long Memory}.
\newblock {\em SSRN eLibrary}, 2001.

\bibitem{LordKahl2006}
Roger Lord and Christian Kahl.
\newblock {Why the Rotation Count Algorithm Works}.
\newblock {\em SSRN eLibrary}, 2006.

\bibitem{melino}
Angelo Melino and Stuart~M. Turnbull.
\newblock Pricing foreign currency options with stochastic volatility.
\newblock {\em Journal of Econometrics}, 45(1-2):239--265, 1990.

\bibitem{muller}
Ulrich~A. Muller, Michel~M. Dacorogna, Rakhal~D. Dave, Richard~B. Olsen,
  Olivier~V. Pictet, and Jacob~E. von Weizsacker.
\newblock Volatilities of different time resolutions -- analyzing the dynamics
  of market components.
\newblock {\em Journal of Empirical Finance}, 4(2-3):213--239, June 1997.

\bibitem{shaw}
William Shaw.
\newblock Stochastic volatility, models of {H}eston type.
\newblock \url{www.mth.kcl.ac.uk/~shaww/web_page/papers/StoVolLecture.pdf}.

\bibitem{zhang}
J.E. Zhang and Jinghong Shu.
\newblock Pricing {S}tandard \& {P}oor's 500 index options with {H}eston's
  model.
\newblock {\em Computational Intelligence for Financial Engineering, 2003.
  Proceedings. 2003 IEEE International Conference on}, pages 85--92, March
  2003.

\end{thebibliography}

\end{document}